\documentclass[journal]{IEEEtran}

\usepackage{siunitx}
\usepackage{amssymb}
\usepackage{bm}
\usepackage{amsmath}
\usepackage{soul}
\usepackage{rotating}
\usepackage{lscape}
\usepackage[export]{adjustbox}
\usepackage{graphics}
\usepackage{cite}
\usepackage{array}
\usepackage{floatrow}
\usepackage{multirow}
\usepackage{graphicx}
\usepackage{caption}
\usepackage{subcaption}
\usepackage[ruled,vlined]{algorithm2e}

\usepackage{enumitem}

\usepackage{authblk}

\usepackage{lipsum}

\ifCLASSINFOpdf
\else
\fi

\begin{document}

% \verso{Krishna Chaitanya \textit{et~al.}}

% \begin{frontmatter}

\title{Semi-supervised Task-driven Data Augmentation for Medical Image Segmentation}
%
%\tnotetext[tnote1]{This is an example for title footnote coding.}

% \author[1]{Krishna \snm{Chaitanya}\corref{cor1}}
% \cortext[cor1]{Corresponding author: 
%   email: krishna.chaitanya@vision.ee.ethz.ch;  
%   Tel: +41 44 63 27908;}
% \author[1]{Neerav \snm{Karani}}
% %\fntext[fn1]{This is author footnote for second author.}
% \author[1]{Christian F. \snm{Baumgartner}}
% %% Third author's email
% %\ead{author3@author.com}
% \author[2]{Anton \snm{Becker}}
% \author[2]{Olivio \snm{Donati}}
% \author[1]{Ender \snm{Konukoglu}}

% \address[1]{Computer Vision Laboratory, ETH Zurich, Sternwartstrasse 7, Zurich-8092, Switzerland.}
% \address[2]{University Hospital of Zurich, Ramistrasse 100, Zurich-8091, Switzerland}

% \received{8 Nov 2019}
% \finalform{}
% \accepted{}
% \availableonline{}
% \communicated{}

\author{\IEEEauthorblockN{Krishna Chaitanya, Neerav Karani, Christian F. Baumgartner, Ertunc Erdil, Anton Becker, Olivio Donati, Ender Konukoglu}
%\thanks{The presented work is partially funded by Swiss Data Science Center (project DeepMicroIA) and the Clinical Research Priority Program Grant on Artificial Intelligence in Oncological Imaging Network from University of Zurich. We also thank NVIDIA corporation for their GPU donation. 
\thanks{KC, NK, CFB, EE and EK are with the Computer Vision Laboratory, ETH Zurich, Switzerland. Email: \{krishna.chaitanya\}@vision.ee.ethz.ch. AB and OD are with the University Hospital of Zurich, Switzerland. \\
Accepted at Medical Image Analysis Journal, 2020.}}

\maketitle

\begin{abstract}
%%%
	Supervised learning-based segmentation methods typically require a large number of annotated training data to generalize well at test time.%to unseen test data.
    In medical applications, curating such datasets is not a favourable option because acquiring a large number of annotated samples from experts is time-consuming and expensive.
    %Curating large labeled datasets in medical applications is not favourable because acquiring a large number of annotated samples from experts is time-consuming and expensive.
    Consequently, numerous methods have been proposed in the literature for learning with limited annotated examples. Unfortunately, the proposed approaches in the literature have not yet yielded significant gains over random data augmentation for image segmentation, where random augmentations themselves do not yield high accuracy. 
    In this work, we propose a novel task-driven data augmentation method for learning with limited labeled data where the synthetic data generator, is optimized for the segmentation task. %which generates augmented data,
	The generator of the proposed method models intensity and shape variations using two sets of transformations, as additive intensity transformations and deformation fields. %model defined in
	Both transformations are optimized using labeled as well as unlabeled examples in a semi-supervised framework. %of the generator model
	Our experiments on three medical datasets, namely cardiac, prostate and pancreas, show that the proposed approach significantly outperforms standard augmentation and semi-supervised approaches for image segmentation in the limited annotation setting. 
	%For cardiac dataset, when only one labeled 3D image is used for training, we get an improvement in mean Dice score of 16.6\% and 18.7\% over standard augmentation and best performing semi-supervised method, respectively.
	The code is made publicly available at https://github.com/krishnabits001/task$\_$driven$\_$data$\_$augmentation.
	%The code is made publicly available~\footnote{https://github.com/krishnabits001/task$\_$driven$\_$data$\_$augmentation}.
%%%%

\end{abstract}

%\begin{keyword}
%\end{keyword}

%\end{frontmatter}

%\linenumbers

% %% main text
% \section{Note}
% \label{sec1}
% Please use \verb+elsarticle.cls+ for typesetting your paper.
% Additionally load the package \verb+medima.sty+ in the preamble using
% the following command: 
% \begin{verbatim} 
%   \usepackage{medima}
% \end{verbatim}

% Following commands are defined for this journal which are not in
% \verb+elsarticle.cls+. 
% \begin{verbatim}
%   \received{}
%   \finalform{}
%   \accepted{}
%   \availableonline{}
%   \communicated{}
% \end{verbatim}

% Any instructions relavant to the \verb+elsarticle.cls+ are applicable
% here as well. See the online instruction available on:
% \makeatletter
% \if@twocolumn
% \begin{verbatim}
%  http://support.stmdocs.com/wiki/
%  index.php?title=Elsarticle.cls
% \end{verbatim}
% \else
% \begin{verbatim}
%  http://support.stmdocs.com/wiki/index.php?title=Elsarticle.cls
% \end{verbatim}
% \fi

\section{Introduction}\label{sec:intro}
% ======================================
% training with a small training dataset is hard
% ======================================
Accurate image segmentation is important for many clinical applications that rely on medical images.
In the recent years, deep neural networks have been successful in yielding high segmentation performance at the expense of requiring large amount of annotated training data.
%In the recent years, deep neural networks have been successful in yielding high segmentation performance. But these methods require large amount of annotated training data to yield such results.
Obtaining many annotated examples is difficult for medical images since getting clinical experts to annotate a large number of segmentation masks, which require per-pixel annotations, is an expensive and time-consuming process.
%Hence, it is not a feasible solution in practical scenarios. 
Hence, it is not a preferable solution in clinical settings.
At the heart of this issue lies the fundamental gap between generalization performance of humans and current Deep Learning (DL) methods. While humans can generalize well for image segmentation after observing very few examples, even one or two examples seem to suffice in some applications, this is not the case with the current DL algorithms. In this work, we focus on algorithmic approaches aiming to close the mentioned gap for medical image segmentation.

In the past works, it has been observed that large annotated data are needed for deep neural networks to obtain high segmentation accuracy. When trained with a large number of annotated samples, the target relationships learned by the neural networks between images and segmentation masks become robust to the variations in shape and intensity characteristics of the target and surrounding structures.
While algorithms trained on a small number of annotated samples, having not been exposed to a sufficient amount of variation, perform poorly on unseen test images that contain variations not observed during training.
%Initial version
% First we contemplate the question: why do we need large annotated data for the deep neural networks to obtain high segmentation accuracy? 
% We hypothesize that for segmentation task, when trained with large number of annotated samples, the target relationship learned by the neural networks between images and segmentation masks is robust to the variations in shape and intensity characteristics of the target and surrounding structures.
% Algorithms trained on a small number of annotated samples may not be exposed to a sufficient amount of variations, and consequently, perform poorly on unseen test images that contain variations not observed during training.
%Variations in shape are due to the tissue properties and its composition, and patient anatomy that differs between individuals. Variations in intensity characteristics are due to the differences in image acquisition and scanner protocols, which are a relevant problem for Magnetic Resonance Imaging (MRI).
These variations in shape arise due to anatomical variations in the population and variations in intensity characteristics are due to differences in (i) tissue properties and its composition, and (ii) image acquisition and scanner protocols, especially in Magnetic Resonance Imaging (MRI). 
%These variations in shape arise due to the difference in patient anatomy between individuals and variations in intensity characteristics are due to (i) difference in tissue properties and its composition, and (ii) differences in image acquisition and scanner protocols, which is a relevant problem for Magnetic Resonance Imaging (MRI). 

To address the need for a large number of annotated data to achieve high segmentation performance with DL methods, in this work we propose a task-driven and semi-supervised data augmentation method (shown in Fig.~\ref{fig:gans}). The method is based on learning generative models that can be used to sample deformation fields and additive intensity transformations. Segmentation cost is included during training of these models for the synthesis of image-label pairs, which incorporate the task-driven nature.
% and the synthesized image-label pairs. 
Semi-supervised nature, on the other hand, is incorporated by including unlabeled data in the training through an adversarial term, which help generators synthesize diverse set of shape and intensity variations present in the population, even in scenarios where the number of labeled examples are extremely low.
%for the case of extremely low annotated examples.
%annotations.
%labeled examples.
%annotations.
%for the case with an extremely low number of labeled examples.
%when the number of labeled examples are extremely low. 

The proposed approach in essence aims to optimize the augmentation task and can be intuitively understood by drawing analogies with existing augmentation methods. 
For example, if we consider random elastic deformations proposed in~\cite{simard2003best,ronneberger2015u}, the augmentation is based on a deformation model with a few number of parameters like Gaussian kernel size and width. These parameters can be seen as hyper-parameters and their values can be optimized by training separate segmentation networks for a number of combinations, and selecting the combination that yields the best performance on validation images. 
Our approach uses a much more flexible transformation model using networks and optimizes its weights using both labeled and unlabeled training images. Its only hyper-parameter is the number of training iterations which is determined based on performance on validation images. Thus, both the training and validation images play a crucial role.
%We perform additional experiments to justify that the proposed approach is not an overfitting scenario for the chosen set of validation images and that it indeed generalizes to any randomly chosen set of images. Surprisingly, we also observe the benefits of the model even for the case when we do not use any validation images in the model selection (training is stopped after a specified number of training iterations).
In our experiments, we evaluated the proposed approach using three different publicly available datasets of cardiac, prostate and pancreas. We present comparisons with existing alternatives as well as an ablation study empirically analyzing the proposed approach. 
%We observed in our evaluation on cardiac, prostate and pancreas dataset that the proposed approach substantially outperforms traditional augmentation and semi-supervised approaches.

A preliminary version of this work has been presented at the conference on Information Processing in Medical Imaging~\cite{kcipmi19}.
In this extended version we additionally: 
\begin{itemize}
    \item analyze quantitatively how each term of regularization loss, namely adversarial loss and large deviation loss components affects the performance gains obtained in the proposed method (refer to results in Sec 5.A).
    \item investigate the benefit of optimizing the generator jointly with the segmentation network as compared to independent optimization of generator w.r.t segmentation network (refer to results in Sec 5.B).
    %\item analyze how each term of regularization in the loss function affects the performance gains of the proposed approach. By this, we quantitatively investigate how each of the adversarial loss and the large deviation loss component affects the performance gains obtained (Refer Results in Sec 5.A).
    %\item explore the effect of independent optimization of the generator w.r.t segmentation network. With this, we evaluate the benefit of optimizing the generator jointly with the segmentation loss as compared to independent training (Refer Results in Sec 5.B).
    \item examine the generality of the proposed method by evaluating it on two more datasets, namely prostate and pancreas.
    %evaluate the proposed method on two more datasets, namely prostate and pancreas to examine its generality.
    \item compare with a larger set of related methods including self-training with conditional random fields and image-level adversarial training.
\end{itemize}
%(i) analyze how each term of regularization in the loss function affects the performance gains of the proposed approach, (ii) explore the effect of independent optimization of the generator w.r.t segmentation network, (iii) evaluate the proposed method on two more datasets namely prostate and pancreas to examine its generality, and (iv) compare with a larger number of methods.
%In this work, we extend the preliminary version in the following directions: (i) we analyze how each term of regularization in the loss function affects the performance gains of the proposed approach, (ii) explore the effect of independent optimization of the generator w.r.t segmentation network, (iii) extensive evaluation of the proposed method was performed on two more datasets namely prostate and pancreas to examine its generality, and (iv) the number of relevant methods compared. % and additional analysis are presented as stated in the sub-sections of the experiments section~\ref{sec:setup}.
%\vspace{-0.4cm}

\subsection{Related work}
We broadly classify the relevant literature into two categories in light of the proposed method in this work:

%%%%%%%%%%%%
%Traditional Data Augmentation
%%%%%%%%%%%%
\textbf{Data Augmentation}: 
Data Augmentation is a simple technique to enlarge the training set based on generating synthetic image-label pairs. 
The idea is to transform the images in such a way that the label remains the same, or the transformation is well-defined for both the image and the label.
The popular approaches are random affine transformations~\cite{cirecsan2011high}, random elastic transformations~\cite{simard2003best,ronneberger2015u} and random contrast transformations~\cite{hong2017convolutional,perez2018data}. These methods are very simple to implement and empirically have been shown to reduce overfitting and improve performance on unseen examples.
Several recent works trained Generative Adversarial Networks(GANs)~\cite{goodfellow2014generative}, using the existing labeled dataset to generate realistic image-label pairs.
The idea has been applied to various analysis tasks~\cite{salehinejad2018generalization,frid2018synthetic,guibas2017synthetic,hou2017unsupervised,wu2018conditional,liu2018pixel,sixt2018rendergan} including MR image segmentation~\cite{costa2018end,shin2018medical,bowles2018gan}. 
MixUp~\cite{zhang2017mixup} is different from the above approaches in the fashion that the generated data is not realistic in nature. Here, the additional data is obtained by linearly interpolating available images and corresponding labels, respectively. Despite the unrealistic nature, it seems to improve the performance of the neural networks on standard benchmark datasets~\cite{zhang2017mixup} and also on medical image segmentation~\cite{eaton2018improving} in limited annotation setting.
All of the above mentioned methods have parameters that control the generation process. These are either set by experience or learned to generate realistic samples based on the available labeled examples, which itself often requires large number of training samples. None of the methods optimize the parameters with respect to the task performance nor leverage unlabeled data. 
The proposed approach optimizes the parameters of the generator to get the best segmentation performance and leverages unlabeled data during the optimization. 

%Meta-learning data augmentations
The closest work to ours was proposed in~\cite{wang2018low}. 
In a meta-learning setup, the authors proposed a fully supervised approach that incorporated the classification performance in learning the generator so to generate augmented images that are optimal for the task. 
Although it is a few-shot learning method, we still need a large number of different classes samples to train the generator. Here, no modeling assumptions are considered in the generator setup, and the augmented images are synthesized directly.
While we instead incorporate domain knowledge and model the generator to output deformation and intensity transformations to generate the synthetic images. Due to these assumptions of modeling transformations as well as leveraging the unlabeled data in the generation process, we do not need a large number of labeled examples during training. In the proposed method, we can readily obtain the synthetic mask using the generated transformation which cannot be obtained with the method~\cite{wang2018low}. %Additionally, we leverage the unlabeled data in a semi-supervised way to capture the useful statistics from the population in the generation process.
We show in the results (Section~\ref{eff_terms_reg} and Fig.~\ref{fig:vary_terms}) that the semi-supervised framework of our method yields significant improvements over using only the labeled examples with task-specific loss as in~\cite{wang2018low}. %, with the task-specific loss.%, i.e., segmentation loss in this case.
In a concurrent work~\cite{paschali2019manifold}, the authors propose a method primarily applicable to classification tasks, where an exhaustive manifold exploration is done to learn affine geometric transformations to find the augmented samples near the decision boundary of the classes. Later, these augmented samples are used with existing training samples to re-train the network, which led to an increase in the robustness (higher accuracy) of the deep learning models. Here, only geometric variations are explored compared to the proposed work where both geometric and intensity variations are addressed. 
Also, instead of optimizing for the task performance as in our model, here authors use augmentation to reinforce the decision boundaries by generating samples near decision boundaries.

%%%%%%%%%%%%
%Semi-supervised learning (SSL)
%%%%%%%%%%%%
\textbf{Semi-supervised learning} (SSL) methods leverage unlabeled data to accompany limited annotated data during training. 
The underlying idea is to regularize training by leveraging the unlabeled data and avoid overfitting.
We divide SSL works into 3 sub-categories: (i) self-training, (ii) adversarial training, and (iii) learned registration based approaches.

%1. Self-training
\textit{Self-training} approaches~\cite{yarowsky1995unsupervised,li2005setred} are based on the concept of iteratively re-training an already-trained network on the estimated labels (pseudo-labels) of unlabeled data as ground truths.
In~\cite{bai2017semi} authors show an improvement in the segmentation performance on cardiac MRIs using a self-training approach along with a Conditional Random Field (CRF) model post-processing intermediate predictions before re-training. 
However, it has also been shown that the self-training approaches can suffer if the initial predictions are erroneous on the unlabeled data~\cite{chapelle2009semi,zhu2009introduction}. 

%2. Adversarial training
\textit{Adversarial} training and GANs have been used in the semi-supervised setting both using the generator as the segmentation network and the discriminator in a regularization term \cite{zhang2017deep}, and vice-versa \cite{souly2017semi}.
In limited annotation setting, we compare and illustrate that the proposed method outperforms the earlier stated semi-supervised approaches.

%4. Adrian's work
Alternatively, in a concurrent work~\cite{zhao2019data}, the authors proposed a one-shot data augmentation approach that learns registration between unlabeled image and labeled image where two independent models are trained to learn spatial and appearance transformations for the registration. Later, both learned models are used to generate augmented image-label pairs which are used to train a segmentation network.
As in the other augmentation works, the generator of this model is also not optimized to yield the best task performance. Furthermore, the approach relies on image registration, which, on one hand, is a difficult task for non-brain anatomy, and on the other hand, leads to task-irrelevant background structures substantially influence the augmentation process.   

%%%%%%%%%%%%
%Weakly-supervised learning
%%%%%%%%%%%%
Lastly, with \textbf{weakly-supervised learning} the issue of expensive and time-consuming pixel-wise annotations is addressed using weaker labels during training such as scribbles~\cite{can2018scribble} and image-wide labels~\cite{andermatt2018pathology}, which are different approaches that can be complementary to data-augmentation. 
%\vspace{-0.1cm}

\section{Methods}
Let $X_L$ be a set of training images and $Y_L$ be the set of corresponding ground truth segmentation labels, $S$ be a segmentation network and $w_s$ be its trainable parameters. In the supervised learning setting, a loss function $L_s \left(X_L,Y_L\right)$ is defined as the disagreement between the labels predicted by the network $S$ for the set of input images $X_L$ and ground truth labels $Y_L$. The objective is to minimize the loss function $L_s$ with respect to the parameters $w_s$ as can be stated as in Eq.~\ref{eq:sup_obj}. 
\begin{equation}
    \min_{w_{S}} L_S(X_L,Y_L)
    \label{eq:sup_obj}
\end{equation}

When data augmentation is used, the minimization becomes% as in Eq.~\ref{eq:sup_plus_aug_obj}. 
\begin{equation}
    \min_{w_{S}} L_S(X_L \cup X_G, Y_L \cup Y_G)
    \label{eq:sup_plus_aug_obj}
\end{equation}
where $X_G$ and $Y_G$ denote the sets of generated images and corresponding labels, which can be generated using the transformations mentioned previously. 
In this minimization, the effective training set is formed of the augmented sets $X_L \cup X_G$ and $Y_L \cup Y_G$. 
When model-based transformations are used for generating the augmentation sets $X_G$ and $Y_G$, such as geometric or contrast transformations, each augmented image $x_G\in X_G$ and the corresponding label $y_G\in Y_G$ are created by a conditional generator function that takes as input an existing labeled pair and applies a random transformation with fixed parameters. We can represent this with the following notation, as in Eq.~\ref{eq:notation}.
%\begin{eqnarray*}
\begin{equation}
\begin{split}
& (x_G, y_G) = G\left((x_L, y_L), z; w_G\right),\\
& (x_L, y_L) \sim p\left(X_L,Y_L\right),\ z\sim p(z),
\label{eq:notation}
\end{split}
\end{equation}    
%\end{eqnarray*}
where $(x_L, y_L)$ are the image-label pair sampled from the set of labeled examples, $G(\cdot,\cdot;w_G)$ is the transformation function, $z$ is the random component of the transformation and $w_G$ are the parameters of the transformation. For instance, for the random elastic deformations proposed in~\cite{ronneberger2015u}, $G$ would be the deformation model, $w_G$ would include the grid spacing between anchor points and standard deviation of the distribution of displacement vectors at each anchor point, while $z$ would correspond to a random draw of displacement vectors. The same transformation would be applied to both $x_L$ and $y_L$ to generate the augmented pairs. As described in the introduction, in the model-based transformations, parameters are often pre-defined and their number is kept low.

When GANs are used for generation, a neural network is used as the generator as $(x_G, y_G) = G(z; w_G)$, and its parameters are determined by optimizing Eq.~\ref{eq:gan} as per the adversarial learning framework~\cite{goodfellow2014generative}, using an additional discriminator network $D$ with its own set of parameters $w_D$. (The GAN can also be a conditional image generator defined as $G(z, x_L; w_G)$ and the discussion still holds.)
\begin{equation}
\begin{split}
    \min_{w_G} \max_{w_D} \mathbb{E}_{x,y\sim p(X_L,Y_L)}[\log D(x, y; w_D)] +\\
    \mathbb{E}_{z\sim p(z)}[\log(1 - D(x_G, y_G; w_D))],\ (x_G, y_G) = G(z; w_G) 
    \label{eq:gan}
\end{split}
\end{equation}
The generator $G$ is input only with a random draw from the distribution $p(z)$ to output an image-label pair $(x_G,y_G)$. The labeled pairs are still used but during the training. The discriminator $D$ is optimized to distinguish between generated pairs from $G$ and real pairs $(x_L,y_L)$, while the generator $G$ is optimized to produce $(x_G,y_G)$ such that $D$ can not differentiate between generated and real pairs. This forces the generated set $(X_G,Y_G)$ to be as "realistic" as possible.

In both model-based and GAN-based approaches, generators would be pre-defined or trained in advanced without considering the task, and during training random draws would be sampled from the generator models to create $X_G$ and $Y_G$.

% ===================================================
% Proposed approach
% ===================================================
\subsection{Semi-Supervised and Task-Driven Data Augmentation}
In this work, we propose to generate augmented image-label pairs that are optimized for the segmentation task (Fig.~\ref{fig:gans}) and the method is summarized in Algorithm~\ref{algo:main_algo}.
We achieve this by optimizing the generator function, similar to the GAN-based approach, but with a crucial difference, we integrate the task loss and leverage unlabeled images in the process. 
The proposed model optimizes Eq.~\ref{eq:sup_plus_aug_task_driven_obj} instead of Eq.~\ref{eq:gan}.
\begin{equation}
    \min_{w_G}\big(\min_{w_{S}} L_S\left(X_L\cup X_G, Y_L\cup Y_G\right) + L_{\text{reg}, G}(X_{UL})\big),
    \label{eq:sup_plus_aug_task_driven_obj}
\end{equation}
where $X_{UL}$ denotes a set of unlabeled images. 
Furthermore, we would like to be able to optimize the parameters with limited number of labeled examples. To this end, we integrate domain knowledge into the generation process, similar to the model-based approaches, but allowing a network parameterization of the transformation models to increase flexibility while remaining trainable. 
We define two conditional generators to model shape and intensity variations using \emph{deformation fields generator} (non-affine spatial transformations) and \emph{intensity fields generator}, respectively. 
%We define two conditional generators to model two factors of variation of shape and intensity using \emph{deformation fields generator} (non-affine spatial transformations) and \emph{intensity transformations generator}, respectively. 
Crucially, both models are constructed such that the segmentation mask $y_{G}$ corresponding to an augmented image $x_{G}$ is obtained by applying the same transformation to the input image mask $y_{L}$.

With this optimization, we want to incorporate two sets of ideas with the two loss-terms in Eq.~\ref{eq:sup_plus_aug_task_driven_obj}. The first term ensures that the model generates set of augmented pairs $(X_G,Y_G)$ such that they are helpful for the minimization of segmentation loss, which is the \emph{task-driven} nature of the approach. 
The second term is a regularization term built on adversarial loss and integrates
a preference for larger transformations leveraging the unlabeled images in the generation process, which is the \emph{semi-supervised} component of the method. 
Also, when using only segmentation loss (task-driven component), generators' may produce images that are easy to segment. Hence, to prevent this case, it becomes crucial to use the regularization loss that comprises of adversarial loss and large deformations loss. Adding large deformations loss and adversarial loss ensures that the deformations and intensity fields generated are larger, and the generated images match the distribution of the unlabeled set. This regularization ensures that the generated images have large changes compared to the input labeled image while also being realistic yet non-trivial to segment for segmentation network.

% ===================================================
% Deformation field generator
% ===================================================
%\vspace{0.1cm}
\subsubsection{Deformation Field Generator}
The deformation field generator $G_V$ is trained to output a deformation field transformation.
The conditional generator $G_V$, a network with parameters $w_{G_V}$, takes as input a labeled image $x_L$ and a $z$ vector randomly drawn from a unit Gaussian distribution to produce a dense per-pixel deformation field $\textbf{v} = G_V(x_L, z; w_{G_V})$. Later, the input image and its corresponding label (in one-hot encoding form) are warped using bi-linear interpolation based on the generated deformation field $\textbf{v}$ to produce the augmented image-label pair $x_{G_V}$ and $y_{G_V}$, respectively. The augmented image and label sets are denoted by $X_{G_V}$ and $Y_{G_V}$, and each sample pair is defined using the following: $x_{G_V}= \textbf{v} \circ x_L$, $y_{G_V}= \textbf{v} \circ y_L$, where $\circ$ denotes warping operation.

% ===================================================
% Additive intensity field generator
% ===================================================
%\vspace{0.1cm}
\subsubsection{Additive Intensity Field Generator}
Similar to $G_V$, here the generator $G_I$ is trained to output an additive intensity mask transformation.
%a diverse set of additive intensity masks such that the augmented images generated from the set \{$X_L$\} can include the distribution of intensity statistics present in the whole population of $\{X_L\} \cup \{X_{UL}\}$.
%intensity masks that can learn the map images from the set \{$X_L$\} to the set of $\{X_L\} \cup \{X_{UL}\}$.
The generator $G_I$, again a network with parameters $w_{G_I}$, takes as input a labeled image from $x_L$ and a $z$ vector randomly drawn from a unit Gaussian distribution to output an additive intensity mask $\Delta I = G_I(x_L, z; w_{G_I})$. Then $\Delta I$ is added to the input image $x_L$ to obtain the augmented image $x_{G_I}$, and its corresponding segmentation mask $y_{G_I}$ remains unchanged as the initial mask $y_L$. The augmented image-label set is denoted by $\{X_{G_I},Y_{G_I}\}$ and one sample pair is defined using the following: $x_{G_I}= x_L + \Delta I $, $y_{G_I}= y_L$.
\setlength{\textfloatsep}{-0.1pt}
\begin{figure}
    \centering
    \begin{subfigure}[b]{0.9\textwidth}
        \includegraphics[width=\textwidth]{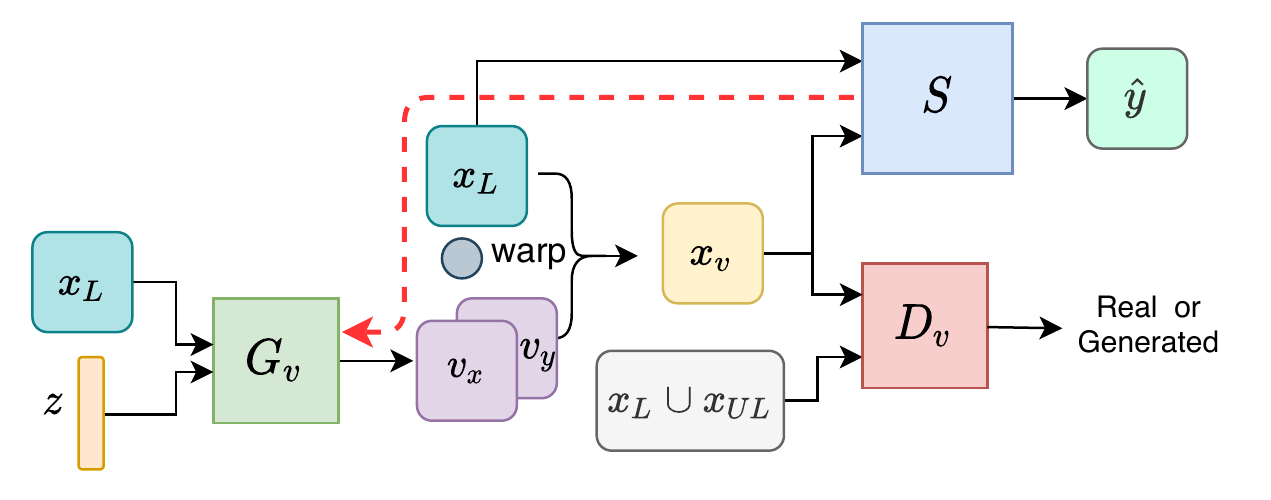}
        \caption{Deformation Field conditional GAN}
        \label{fig:df_gan}
    \end{subfigure}
    \begin{subfigure}[b]{0.9\textwidth}
        \includegraphics[width=\textwidth]{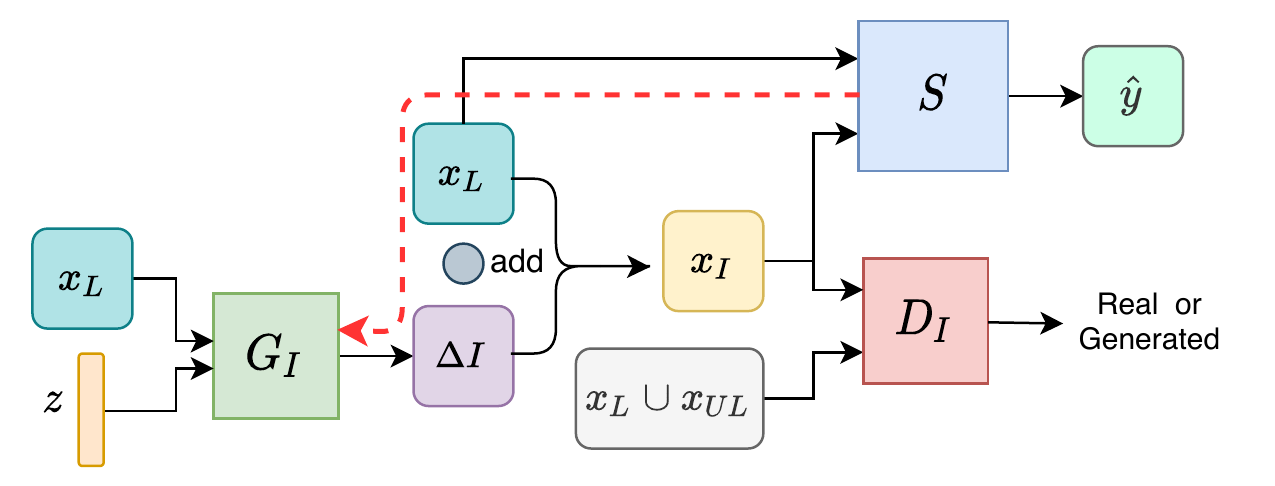}
        \caption{Additive Intensity Field conditional GAN}
        \label{fig:ci_gan}
    \end{subfigure}
    \caption{Data augmentation modules that generate augmented image-label pair with task-driven optimization defined in a semi-supervised framework. Here, the dotted-red line indicates the inclusion of segmentation loss for generator optimization.}
    %\vspace{-0.25cm}
    \label{fig:gans}
\end{figure}
% ===================================================
% Regularization term
% ===================================================
\subsubsection{Regularization Loss}
The regularization term $L_{\text{reg}}$ is defined as in Eq.~\ref{eq:gan_G_C} for both conditional generators.
\begin{equation}
    \label{eq:gan_G_C} L_{\text{reg}, G_C} = \lambda_{\text{adv}}L_{\text{adv}, G_C} + \lambda_{\text{LD}}L_{\text{LD}, G_C},\ \text{for }C={V,I}.
\end{equation}
The first term is the following standard adversarial loss
%\begin{equation}\label{eq:adv_G_C}
%\begin{split}
%    & L_{\text{adv}, G_C} = \\
%    & \min_{w_{G_{C}}} \mathbb{E}_{z\sim p(z),x_{L}\sim p(X_L)}[\log(1 - %D_{C}(G_C(x_L, z; w_{G_C}); w_{D_C}))]
%\end{split}
%\end{equation}
\begin{equation}\label{eq:adv_G_C}
\begin{split}
    {} & L_{\text{adv}, G_C} = \max_{w_{D_{C}}} \mathbb{E}_{x\sim p(X_{U} \cup X_{UL})}[\log D_{C}(x; w_{D_C})] + \\
    & \mathbb{E}_{z\sim p(z),x_{L}\sim p(X_L)}[\log(1 - D_{C}(G_C(x_L, z; w_{G_C}); w_{D_C}))]
\end{split}
\end{equation}
where $D_C$, $w_{D_C}$ denotes the discriminator network and its weights. 
This first term incorporates the semi-supervised nature of the model by including the set of unlabeled images $X_{UL}$ in the optimization. 
It allows samples in the unlabeled set that show different shape and intensity variations than the labeled examples guide the optimization of the generators. 
The second term in Eq.~\ref{eq:gan_G_C} is the \emph{Large Deviation} (LD) term that embeds a preference for large transformations. 
The LD term prevents the generator from producing relatively smaller deformations and intensity fields, which would satisfy the adversarial loss as well as lead to lower segmentation loss in the cost given in Eq.~\ref{eq:sup_plus_aug_task_driven_obj}. 
The definition of $L_{\text{LD},G_C}$ depends on the generator type. We define the two terms for the deformation and intensity field generators as $L_{\text{LD}, G_V} = -\|\textbf{v}\|_1$ and $L_{\text{LD}, G_I} = -\|\Delta_{I}\|_1$. The negative signs ensures that minimizing the LD term maximizes the $L_1$ norms. 
The LD term forces the generator to produce large transformations while the adversarial loss tries to constrain them. 
Optimizing for both terms yield augmented images that differ substantially from the input labeled images for both the generators. 

Finally, the weighting terms $\lambda_{adv}$ and $\lambda_{\text{LD}}$ balances the effect of the two terms on the optimization. 

\textit{Discriminator:}
The discriminator's role is to ensure that the augmented images obtained by applying transformations produced by the generators are indistinguishable from real unlabeled images in the distribution sense. This loss competes against segmentation loss to ensure that the generated images are relatively realistic and are not too easy to segment images either.

% ===================================================
% Optimization
% ===================================================
%\vspace{0.1cm}
\subsubsection{Optimization Sequence}
Before the optimization of the conditional generators, the segmentation network ($S$) in the proposed setup is pre-trained on only the labeled data for a few epochs as per Eq.~\ref{eq:sup_obj} and later optimized with both the labeled and generated data from the conditional generators as per Eq.~\ref{eq:sup_plus_aug_obj}. This is done to ensure that $S$ produces reasonable segmentation masks for the labeled data when the optimization of the generators begins.
Following the pre-training of $S$, both generative models for deformation and intensity fields are trained separately by minimizing the cost given in Eq.~\ref{eq:sup_plus_aug_task_driven_obj} with corresponding regularization terms. Note that this minimization trains over $S$, the generators and the discriminators. This implies training of the set of networks $(S,G_V,D_V)$ and $(S,G_I,D_I)$ independently for deformation and intensity fields generation, respectively.
The optimization sequence is run for a fixed number of iterations, and the segmentation network is evaluated on the validation images using the Dice's similarity coefficient (DSC) at every iteration. 
Once the optimization is complete, we fix both the generators $G_V$ and $G_I$ with the parameters that yielded the best mean DSC over the validation images.
Then, the segmentation network $S$ is optimized using Eq.~\ref{eq:sup_plus_aug_obj} once again from a random initialization. 
The data used for this final training comprises of both the original labeled sets, $X_L$ and $Y_L$, and the augmented sets, $X_G$ and $Y_G$ sampled from the trained generators $G_V$ or $G_I$ or both (where we perform two back to back transformations, e.g. $G_I$ is applied over the deformed image obtained from $G_V$).

The validation images play a crucial role in the defined optimization as they determine the parameters of the generator model chosen to generate the augmented data that is later used for the independent optimization of the segmentation network.
%We stress the importance of the validation images in the optimization of the parameters of the generative model that is used for augmentation. During the optimization, the selected model is the one that yields the best Dice score on the validation images over all the training iterations. 
%\vspace{-0.3cm}

%\setlength{\textfloatsep}{-0.05pt}
\begin{algorithm}[!ht]
\SetAlgoLined
Available training data: labeled set $(X_{L},Y_{L})$ and unlabeled set $(X_{UL})$.\\

\textbf{Step 1}:\\
(a) train deformation field generator $(G_{V})$ as per Eq. (5) using the available data.\\
(b) train intensity field generator $(G_{I})$ as per Eq. (5) using the available data.\\

\textbf{Step 2}:\\
Post optimization, the generators are used to sample augmented image, label pairs conditioned on the labeled set. \\
(a) The shape transformed image, label pairs $(X_{G_V}, Y_{G_V})$ are sampled using generator $(G_{V})$.\\
(b) The intensity transformed image, label pairs $(X_{G_I}, Y_{G_I})$ are sampled using generator $(G_{I})$.\\
(c) The image, label pairs that contain both the shape and intensity transformations (denoted by $(X_{G_{VI}}, Y_{G_{VI}})$) are obtained by inputting the sampled shape transformed image, label pairs $(X_{G_V}, Y_{G_V})$ from $(G_{V})$ through $(G_{I})$.\\
 
\textbf{Step 3}:\\
Train the segmentation network with all the available labeled and generated augmented data.
%Train the segmentation network with all the available labeled data ($(X_{L}, Y_{L})$, affine transformed data $(X_{Aff}, Y_{Aff})$) and generated augmented data (shape transformed data $(X_{G_V}, Y_{G_V})$, intensity transformed data $(X_{G_I}, Y_{G_I})$ and data with both shape and intensity transformations applied $(X_{G_{VI}}, Y_{G_{VI}})$).
The training set includes: original labeled data $(X_{L}, Y_{L})$, affine transformed data $(X_{Aff}, Y_{Aff})$, shape transformed data $(X_{G_V}, Y_{G_V})$, intensity transformed data $(X_{G_I}, Y_{G_I})$ and data with both shape and intensity transformations applied $(X_{G_{VI}}, Y_{G_{VI}})$. 

\caption{Training steps of the proposed method}
\label{algo:main_algo}
\end{algorithm}

\section{Dataset and Network details}
\subsection{Dataset details}
\textbf{Cardiac Dataset}\label{sec:dataset_acdc}: This is a publicly available dataset hosted as part of MICCAI'17 ACDC challenge~\cite{bernard2018deep}~\footnote{https://www.creatis.insa-lyon.fr/Challenge/acdc}. 
It comprises of 100 subjects' short-axis cardiac cine-MR images captured using 1.5T and 3T scanners. The in-plane resolution ranges from 0.70x0.70mm to 1.92x1.92mm and through-plane resolution ranges from 5mm to 10mm. The segmentation masks are provided for left ventricle (LV), myocardium (Myo) and right ventricle (RV) for both end-systole (ES) and end-diastole (ED) phases of each subject. This dataset is divided into 5 sub-groups (details in ~\cite{bernard2018deep}~\footnote{https://www.creatis.insa-lyon.fr/Challenge/acdc}), each comprising of 20 subjects, respectively. 
%These 5 sub-groups are abnormal right ventricles, dilated cardiomyopathy, normal controls, previous myocardial infarction and hypertrophic cardiomyopathy.
For our experiments in the main article, we only used the ES images, which was a random choice.
The evaluation of ED images for the proposed method are presented in table~\ref{table:acdc_ed_images} in the appendix.

\textbf{Prostate Dataset}: This is a public dataset made available as part of MICCAI'18 medical segmentation decathlon challenge~\footnote{http://medicaldecathlon.com/index.html}. It comprises of 48 subjects T2 weighted MR scans of prostate. The in-plane resolution ranges from 0.60x0.60mm to 0.75x0.75mm and through-plane resolution ranges from 2.99mm to 4mm. Segmentation masks comprise of two adjoint regions: peripheral zone (PZ) and central gland (CG). 

\textbf{Pancreas Dataset}: This dataset also is from medical decathlon MICCAI'18 challenge~\footnote{http://medicaldecathlon.com/index.html}. It comprises of 282 subjects CT scans.
Segmentation masks available comprise of labels with large (background), medium (pancreas) and small (tumor) structures. The in-plane resolution ranges from 0.6x0.6mm to 0.97x0.97mm and through-plane resolution ranges from 0.7mm to 7.5mm. 
In this work, we create two labels for segmentation task, where label 1 denotes the foreground which was created by merging the labels of pancreas and tumor, and label 0 denotes the background label.
\vspace{-0.25cm}

\subsection{Pre-processing}
N4~\cite{tustison2010n4itk} bias correction was performed on the cardiac and prostate datasets.
The below pre-processing was applied to all images of all the datasets.
(i) intensity normalization: all the volumes were normalized using min-max normalization according to: $(x-x_2)/(x_{98}-x_2)$, where $x_2$ and $x_{98}$ denote the $2^{nd}$ and $98^{th}$ intensity percentiles of the 3D volume.
(ii) re-sampling: all 2D image slices and their corresponding label maps were re-sampled to a fixed in-plane resolution $r$ using bi-linear and nearest-neighbour interpolation, respectively and then cropped or padded with zeros to a fixed size of $f_z$. 
The resolution $r$ and fixed size $f_z$ were chosen empirically for each dataset, and the values were:
(a) cardiac dataset: $r = $1.367x1.367mm and $f_z = $224x224, 
(b) prostate dataset: $r = $0.6x0.6mm and $f_z = $224x224, 
(c) pancreas dataset: $r = $0.8x0.8mm and $f_z = $320x320.
%\vspace{-0.25cm}

\subsection{Network Architecture}
There are 3 types of networks in the proposed method (see Fig.~\ref{fig:gans}): a segmentation network $S$, a discriminator network $D$ and a generator network $G$.
We describe the architectures of these networks below. $G_V$ and $G_I$ use the same architecture except at the last layer, which are used to model the deformation field and the intensity mask, respectively.

\textbf{Generator}: 
Generator $G$ takes an image $x_L$ and a randomly drawn $z$ vector of dimension 100 as input. Both inputs are initially passed through 2 sub-networks namely $G_{subnet,X}$ and $G_{subnet,z}$. $G_{subnet,X}$ comprises of 2 convolutional layers and $G_{subnet,z}$ comprises of a fully-connected layer, followed by reshaping of output into down-sampled image dimensions. Then a set of bi-linear upsampling and convolutional layers are applied consecutively to output feature maps of image dimensions.
The resulting outputs of both the sub-networks, which are of same dimensions, are then concatenated and passed through a common sub-network $G_{subnet,common}$, which consists of 4 convolutional layers where the last layer is different for $G_V$ and $G_I$. The final convolution for $G_V$ yields two feature maps that correspond to dense per-pixel deformation field $\textbf{v}$, while for $G_I$, it outputs a single feature map that corresponds to the intensity mask $\Delta I$.
The final layer of $G_I$ uses $tanh$ activation to restrict the range of values in the intensity mask. All the convolutional layers except the final layers are followed by batch normalization layers and ReLU activation.
All convolutional layers' kernels are 3x3 except the final layers' which are 1x1.

\textbf{Discriminator}:
$D$ has an architecture similar to the DCGAN~\cite{radford2015unsupervised}, which comprises of five convolutional layers each with a kernel size of 5x5 and a stride of 2. The convolutions are followed by batch normalization layers and leaky ReLU activations with the leak value of the negative slope set to 0.2.
After the convolutional layers, the output is reshaped and passed through three fully-connected layers where the final layer has an output size of 2 that predicts the probability of the input being real or fake.

\textbf{Segmentation Network}:
We chose the architecture of segmentation network $S$ similar to U-Net~\cite{ronneberger2015u}. It comprises of encoding and decoding paths. The encoder comprises of four convolution blocks where each block has two 3x3 convolutions followed by a 2x2 max-pool layer. The decoder comprises of four convolution blocks where each block comprises of the concatenation of features from the corresponding level of the encoder, followed by two 3x3 convolutions and bi-linear upsampling by a factor of 2. Except for the last layer, all the layers have batch normalization and ReLU activation.
%\vspace{-0.25cm}

\subsection{Training Details}
The segmentation loss ($L_S$) used is weighted cross entropy.
We empirically set the weights of background pixels as 0.1 and foreground pixels as 0.9 since the number of pixels belonging to the foreground are fewer in quantity and are of primary interest for the segmentation task at hand. For the datasets with more than one foreground label, we divided the foreground weight of 0.9 equally among all the labels. With this rationale, the weights of the output labels are as follows: (a) 0.1 for background and 0.3 for each of the three foreground classes of the cardiac dataset, (b) 0.1 for background and 0.45 for each of the two foreground classes of the prostate dataset, and (c) 0.1 for background and 0.9 for one foreground class of the pancreas dataset.

We split the data into training pool, validation, unlabeled and test sets.
We empirically set $\lambda_{adv}$ as 1 to match the magnitude of adversarial loss to the segmentation loss.
To determine $\lambda_{LD}$ parameter, we randomly sampled one 3D volume from the training pool of cardiac dataset (the sampled volume is one possibility of all training combinations used in full analysis) and trained the network, and evaluated performance on the validation images for three values of $\lambda_{LD}$: $10^{-2},10^{-3},10^{-4}$.
The value of $10^{-3}$ for $\lambda_{LD}$ yielded the best validation performance for this experiment. So, we used this set values ($\lambda_{adv}=1, \lambda_{LD}=10^{-3}$) for all future experiments on all datasets. 
Owing to the computationally expensive nature of the proposed method, we did not perform an exhaustive grid search on many combinations of weights of the loss terms ($\lambda_{adv}$, $\lambda_{LD}$) on the validation set. But if one has enough computational resources, one could do a grid search of these hyper-parameters for each dataset to potentially obtain higher performance gains. 
The training of the GANs to convergence with limited data with stability can be complicated, as seen in many prior applications. We did not face any major training issues as: (1) we incorporated the concluding remarks from earlier works like DCGAN~\cite{radford2015unsupervised} into the designing of our generator and discriminator networks and choosing appropriate respective learning rates, and also here (2) the generators are trained to generate deformation/intensity fields instead of images directly as done in earlier works, which could be relatively easier to train.

% We empirically set $\lambda_{adv}$ as 1.
% We split the data into training pool, validation, unlabeled and test sets.
% To determine $\lambda_{LD}$ parameter, we randomly sampled one 3D volume from the training pool (the sampled volume is one possibility of all training combinations used in full analysis) and trained the network, and evaluated performance on the validation images for three values of $\lambda_{LD}$: $10^{-2},10^{-3},10^{-4}$.
% The value of $10^{-3}$ for $\lambda_{LD}$ yielded the best validation performance for this experiment for all datasets. So, we used this value for all the future experiments.
% We did not perform a grid search to try many combinations of weights of the loss terms ($\lambda_{adv}$, $\lambda_{LD}$) on the validation set beyond three values for only one factor,i.e., $\lambda_{LD}$. But one can do a grid search of these hyper-parameters for each dataset to obtain higher performance gains.
%%%Below values yielded the best validation performance in this experiment for each dataset and subsequently used for all the experiments: (a) $10^{-3}$ for cardiac dataset, $10^{-2}$ for prostate dataset and (c) $10^{-2}$ for pancreas dataset.
The batch size and the total number of iterations are set as 20 and 10000, respectively, based on the evolution of the training curves.
For all the networks, while training, the iteration where the model yields the best performance on the validation images is saved for the evaluation.
AdamOptimizer~\cite{kingma2014adam} is used for the optimization of all the networks with learning rate of $10^{-3}$ and default beta values ($\beta_1=0.9$, $\beta_2=0.999$).
%\vspace{-0.05cm}

\section{Experiments}\label{sec:setup}
% ================================
% number of images in training / validation / test
% ================================
We evaluated the proposed method on three datasets: cardiac, prostate and pancreas.
For each dataset, we split the data into 4 sets: labeled training ($X_{L,total}$), unlabeled training ($X_{UL}$), test ($X_{ts}$) and validation ($X_{vl}$). 
The size of each set is denoted by $N$ followed by a subscript indicating the set.
The validation set consists of two 3D volumes ($N_{vl}$=2) for all datasets. For the cardiac, prostate and pancreas datasets, the number of 3D volumes ($N_{UL}$, $N_{ts}$) for unlabeled and test sets are (25, 20), (20, 13) and (25, 20) respectively.
$X_{UL}$, $X_{ts}$ and $X_{vl}$ sets are selected randomly a-priori and fixed for all experiments. 
The remaining 3D volumes constitute the training pool $X_{L,total}$.
Note that the entire training pool is never utilized for training. Rather a small number of training images ($N_{L}$) is sampled from $X_{L,total}$ for each experiment.
As the interest of this work is to analyze the performance in the limited annotation setting, we set $N_{L}=1$ or $3$.
Each experiment is run five times with different 3D training volumes. Further, to account for the variations in the random initialization and convergence of the networks, each of the five experiments is run three times.
Thus, overall, we have 15 runs for each experiment.
Since the cardiac dataset has five sub-groups of images (see Sec.~\ref{sec:dataset_acdc}), we ensure that each set contains an equal number of images from each sub-group, and, when $N_{L}=1$ is run five times, each time the 3D volume is selected from a different sub-group.

% ================================
% Experiment setup 
% ================================
%\vspace{0.1cm} 
%\noindent The following models were evaluated and compared in segmentation accuracy:
\noindent Segmentation performances of the following models were compared:
% \begin{itemize}
%     %1,1 No data augmentation 
%     \item \textbf{No data augmentation (Aug\textsubscript{none})}: $S$ is trained without any data augmentation.
%     %1,2 Affine data augmentation 
%     \item \textbf{Affine data augmentation(Aug\textsubscript{Aff})}: $S$ is trained with data augmentation comprising of affine transformations.
%     The transformations comprised of: (a) scaling (random scale factor is chosen from a uniform distribution with min and max value as 0.9 and 1.1), (b) flipping along x-axis, (c) rotation (randomly a value is chosen between -15 and +15 degrees and another type of rotation that is multiple of 45 degrees (defined as 45 deg*N where N is randomly chosen between 0 and 8)).
%     For each slice in the batch, we apply one of the above random transformation 80\% of the time and 20\% of the time we use the image as is. 
%     %For each slice in a batch, we sample a random number between 0 and 5. If the number sampled is between 0 and 3, we apply the specific augmentation while for 4 we use the default slice and do not apply any transformation.
% \end{itemize}

%1,1 No data augmentation Aug\textsubscript{none}
\vspace{0.1cm} \textbf{No data augmentation (no aug)}: $S$ is trained without any data augmentation.

%1,2 Affine data augmentation Aug\textsubscript{Aff}
\vspace{0.1cm} \textbf{Affine data augmentation (Aff)}: $S$ is trained with data augmentation comprising of affine transformations such as: (a) scaling (random scale factor is chosen from a uniform distribution with min and max value as 0.9 and 1.1), (b) flipping along x-axis, (c) rotation (randomly a value is chosen between -15 and +15 degrees and another type of rotation that is multiple of 45 degrees (defined as 45 deg*N where N is randomly chosen between 0 and 8)).
For each slice in the batch, we apply one of the above random transformation 80\% of the time and 20\% of the time we use the image as is. 

%\vspace{0.1cm}     
%\noindent 
For all the subsequent data augmentation methods, MixUp and semi-supervised methods, we include the random affine data augmentations by default as described above. For training of $S$, half of each batch was composed random affine augmentation and the remaining half was chosen from the specific augmentation technique. 
\vspace{0.1cm} \textbf{Random elastic deformations (RD)}:
Random elastic augmented images are created as stated in~\cite{ronneberger2015u}, where a deformation field is created using a matrix of size 3x3x2. Each element of this matrix is sampled from a Gaussian distribution with a mean of 0 and standard deviation (sigma) of 10 and is then re-sampled to image dimensions using bi-cubic interpolation. 
(refer to Fig.~\ref{fig:rd_vary_sig_ks} in the appendix for results with different sigma \& matrix sizes)

%3.2 Random contrast and brightness augmentation Aug\textsubscript{RI}
\vspace{0.1cm} \textbf{Random contrast and brightness fluctuations}~\cite{hong2017convolutional,perez2018data} \textbf{(RI)}:
These augmented images are created with the help of contrast adjustment step ($x = (x - \Bar{x}) * c + \Bar{x}$) and brightness adjustment step ($x = x + b$), where c and b are sampled uniformly from [0.8,1.2] and [-0.1,0.1], respectively and $\Bar{x}$ denotes mean of 2D image. 
(refer to Fig.~\ref{fig:rd_vary_cont_brit} in the appendix for more combinations of c and b evaluated)

%4 Deformation field cGAN (Aug\textsubscript{GD})
\vspace{0.1cm} \textbf{Deformation field transformations (GD)}: The deformation field generator trained with the proposed method $G_V$ is used to generate the augmented data i.e., $X_{G_V}$.

%5 Intensity field cGAN Aug\textsubscript{GI}
\vspace{0.1cm} \textbf{Intensity field transformations (GI)}: The intensity field generator trained with the proposed method $G_I$ is used to generate the augmented data i.e., $X_{G_I}$.

%6. Both GANs augmented data Aug\textsubscript{GD+GI}
\vspace{0.1cm} \textbf{Both deformation and intensity field transformations (GD+GI)}: Augmented data included both $X_{G_V}$ and $X_{G_I}$, obtained from the generators $G_V$ and $G_I$, respectively. 
We also generated additional images $X_{G_{VI}}$ which have both the deformation and intensity transformations. These are obtained by applying intensity transformation using generator $G_I$ on the deformation field transformed images $X_{G_V}$. The augmented data consists of all the images generated $\{X_{G_V},X_{G_I},X_{G_{VI}}\}$.

%7. Mix up Aug\textsubscript{Mixup}
\vspace{0.1cm} \textbf{MixUp}~\cite{zhang2017mixup} \textbf{(Mixup)}:
The augmentation sets $X_G$ and $Y_G$ consist of the linear combination of available labeled images using the Mixup formulation as stated in Eq.~\ref{eq:mixup_eqn}~\cite{zhang2017mixup}.
\begin{equation}
    {x_{Gi}} = \lambda x_{Li} + (1-\lambda) x_{Lj},\hspace{0.2cm}
    {y_{Gi}} = \lambda y_{Li} + (1-\lambda) y_{Lj}
    \label{eq:mixup_eqn}
\end{equation}
where $\lambda$ is sampled from beta distribution Beta$(\alpha,\alpha)$ with $\alpha \in (0,\infty)$ and $\lambda \in [0,1)$. The $\alpha$ value of 0.2 yielded the best results for the datasets considered. $\lambda$ controls the ratio of mixing of two image-label pairs $(x_{Li},y_{Li})$, $(x_{Lj},y_{Lj})$ which are randomly sampled from the labeled image set.

%8. Mixup + our method Aug\textsubscript{GD+GI+Mixup}
\vspace{0.1cm} \textbf{Mixup over deformation and intensity field transformations (GD+GI+Mixup)}:
These set of images are obtained by applying Mixup over all possible pairs of available images: original data ($X_L$), their affine transformations and the generated images using deformation and intensity field generators

%9. Semi-supervised Learning 
\vspace{0.1cm} \textbf{Adversarial Training (Adv\_tr)}:
For comparison, we investigate previously proposed adversarial training methods with the discriminator trained to operate on: (i) the image level discrimination~\cite{zhang2017deep} and (ii) the pixel level discrimination~\cite{souly2017semi} in a semi-supervised (SSL) setting. 
%For comparison, we investigate previously proposed adversarial training methods with the discriminator trained to operate on: (i) the image level discrimination~\cite{luc2016semantic} in both supervised (SL)~\cite{luc2016semantic} and semi-supervised (SSL)~\cite{zhang2017deep} setting and (ii) the pixel level discrimination~\cite{souly2017semi} in a semi-supervised (SSL) setting. 

%10. Self-training
\vspace{0.1cm} \textbf{Self-training (Self\_tr)}:
The self-training based method as proposed in~\cite{bai2017semi} is evaluated on the datasets.

\vspace{0.5cm}
\textbf{Ablation Studies:}
In addition to the aforementioned comparisons, we carried out additional ablation studies as described below. These studies were done only on the cardiac dataset owing to lack of computational resources, as each experiment for each ablation study and dataset requires 15 runs. 
%o understand the effects of different terms of regularization 
%below ablation studies are done on cardiac dataset.

%\begin{itemize}
%\begin{enumerate}[label=(\Alph*)]
%11. Varying the terms of regularization
\vspace{0.1cm} \textbf{A. Effects of adversarial ($\lambda_{adv}$) and large deviation ($\lambda_{LD}$) loss terms of regularization loss on segmentation performance}:
We investigate the effect of each term of regularization on the performance of the proposed method. Different values of $\lambda_{adv}$ and $\lambda_{LD}$ are considered to examine how much each term impacts the performance. The training case of $\lambda_{adv} = \lambda_{LD} = 0$ is similar to earlier work~\cite{wang2018low}.
%     %We enable or disable different terms of regularization to analyze the impact on the performance of the proposed method. Different combinations of $\lambda_{adv}$ and $\lambda_{LD}$ are enabled or disabled to analyze how each term impacts the performance.

%12. independent optimization
\vspace{0.1cm} \textbf{B. Independent optimization of the generator and the segmentation networks}:
Here, we optimize both $G_V$ and $G_I$ without the segmentation loss similar to~\cite{shin2018medical,bowles2018gan}. Later, augmented data created from these optimized generators are used for the independent training of the segmentation network. This experiment reveals the value of the segmentation loss.
%Later the optimized generators are used to create augmented data, utilized for the independent training of the segmentation network. This experiment reveals the value of the segmentation loss. %We study how much this dis-joint optimization can affect the segmentation performance.

%13. Varying the number of unlabeled data
\vspace{0.1cm} \textbf{C. Varying the number of unlabeled data}:
We used different number of unlabeled volumes in the training of the generators. The number of 3D volumes studied ($N_{UL}$) were: 1, 3, 5, 10, 20, 25 and 50.
%vary the number of unlabeled 3D volumes used for the training of the generators to analyze the effect on the final performance. The number of 3D volumes studied ($N_{UL}$) were: 5, 10, 20 and 25.

%14. Varying the number of labeled data in training
\vspace{0.1cm} \textbf{D. Varying the number of labeled 3D volumes used in training}:
%We vary the number of labeled 3D volumes used for the training of both the proposed generators to study the effect on the final performance. 
The number of 3D volumes considered ($N_{L}$) were: 1, 3, 5, 10, 15, 40. This experiment is done to study if any improvement in Dice score is obtained when a large number of annotated volumes are available for training. 

\vspace{0.1cm} \textbf{E. Different set of train, validation, test and unlabeled 3D volumes}:
We randomly sample another training, validation, test, and unlabeled image sets from the cardiac dataset different from the earlier sets and re-run learning deformation and intensity field transformations for this new set for one 3D training volume case ($N_{L}=1$). This experiment is done to analyze if the proposed method overfits to a specific dataset split or generalizes for any split.
%. The experiment of learning deformation and intensity field transformations is re-run on the new set for the case of training with one 3D volume ($N_{L}=1$). This experiment is done to analyze if the proposed method overfits to a specific dataset split or generalizes for any split.% for any randomly chosen split to yield the performance gains. 

\vspace{0.1cm} \textbf{F. No validation images}: 
Lastly, we report the performance observed when we do not use any validation images in the training of the generators. In this case, the training is stopped after running the model for some predefined number of iterations and these model parameters are used for generating images for augmentation. 
%\end{itemize}

\noindent \textbf{Evaluation}: 
Dice's similarity coefficient (DSC) is used to evaluate the segmentation performance of each method. The performance reported is obtained on $N_{ts}$ number of test images for the structures of each dataset as stated earlier.
%\vspace{-0.05cm}

\section{Results and Discussion}\label{sec:results}

% ================================================================
% quantitative results - for cardiac, prostate, pancreas
% ================================================================
%\setlength{\belowcaptionskip}{-5pt} [htb!]
\begin{figure}[!t]
    \centering
    \begin{subfigure}{0.9\textwidth}
        \includegraphics[width=\textwidth]{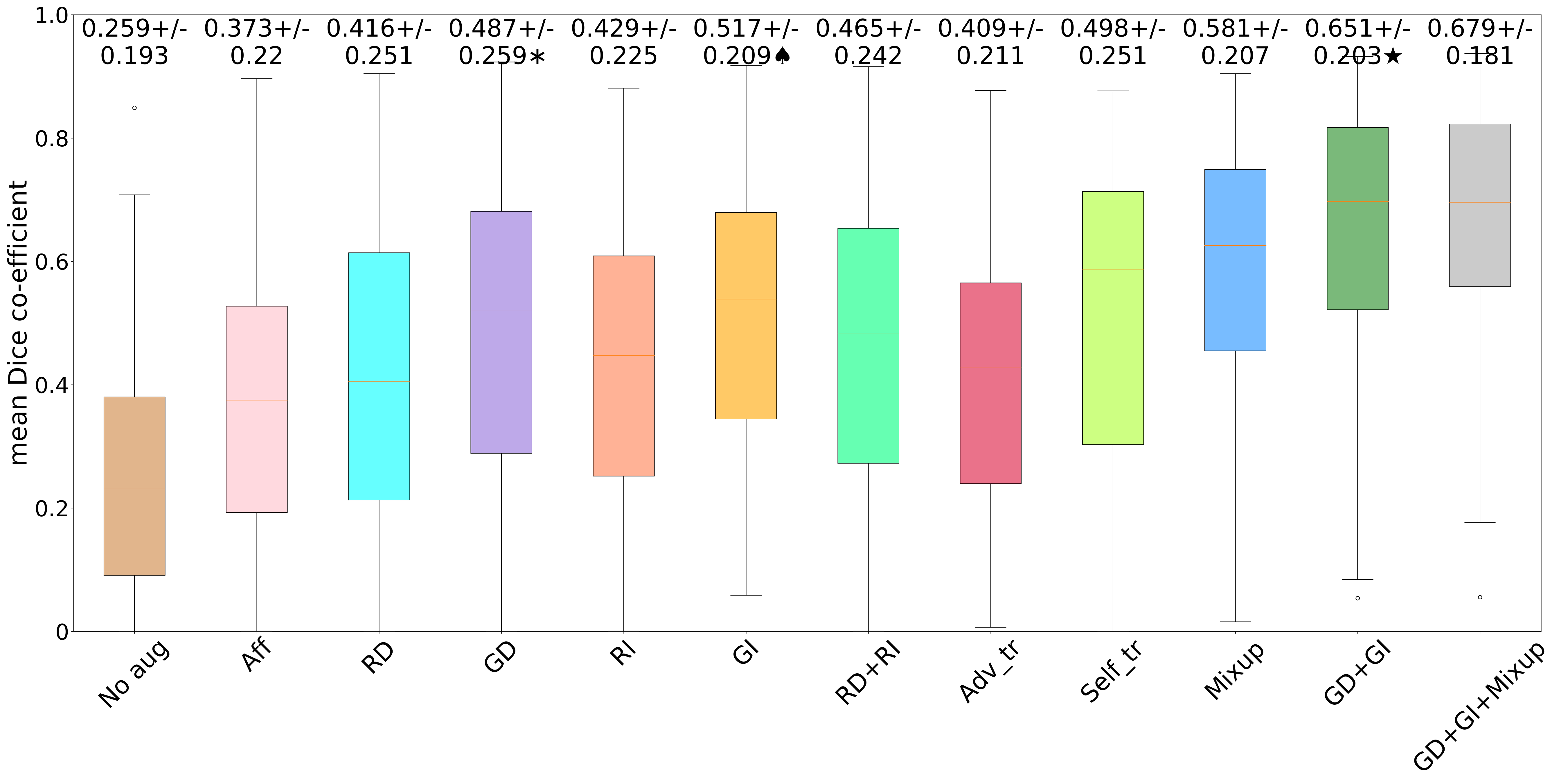}
        \caption{Right Ventricle (Cardiac data)}
        \label{fig:rv_res}
    \end{subfigure}
    \begin{subfigure}{0.9\textwidth}
        \includegraphics[width=\textwidth]{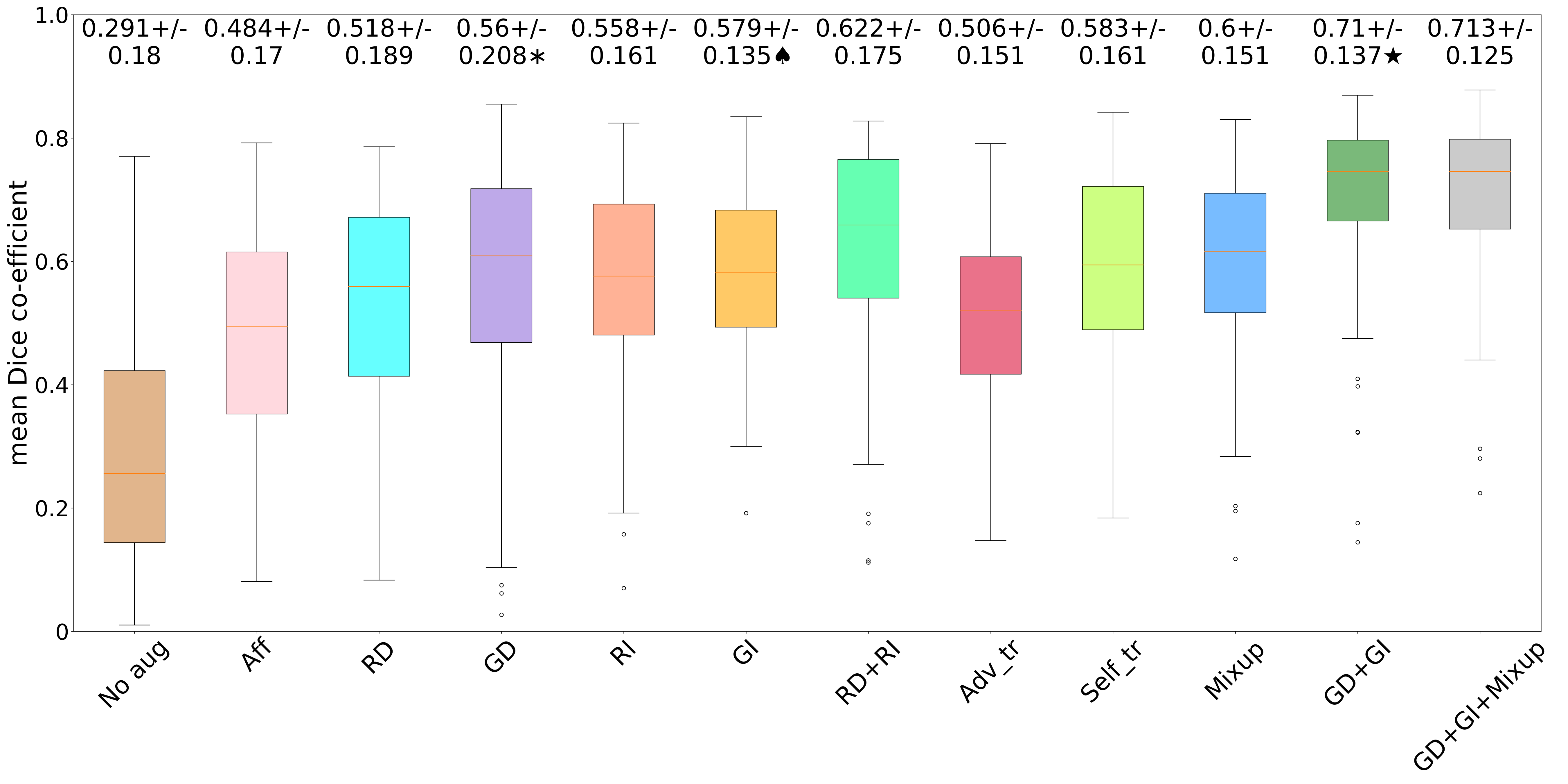}
        \caption{Myocardium (Cardiac data)}
        \label{fig:myo_res}
    \end{subfigure}
    \begin{subfigure}{0.9\textwidth}
        \includegraphics[width=\textwidth]{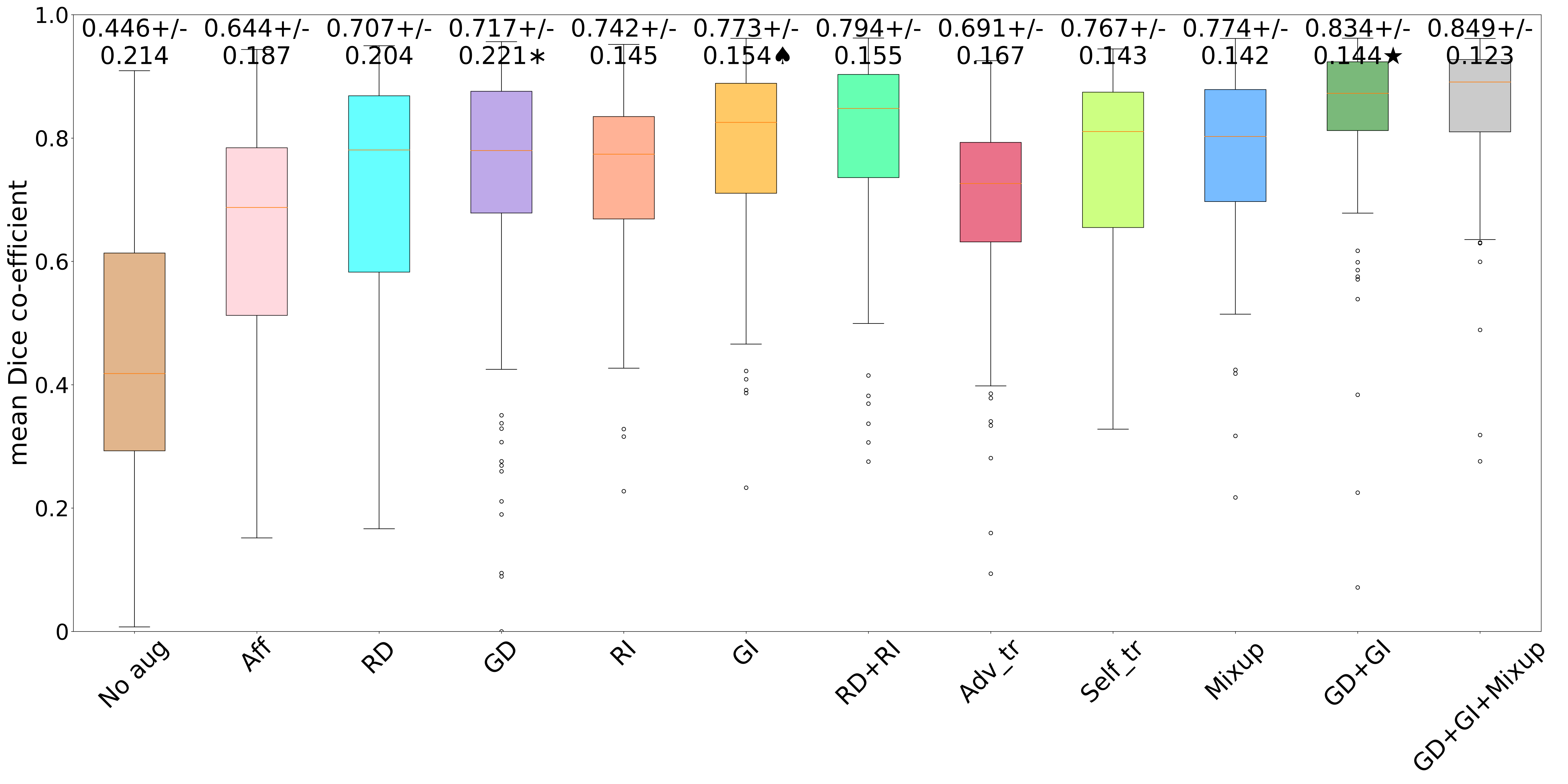}
        \caption{Left Ventricle (Cardiac data)}
        \label{fig:lv_res}
    \end{subfigure}
\end{figure}
\begin{figure}[!ht]\ContinuedFloat
    \centering
    \begin{subfigure}{0.9\textwidth}
        \includegraphics[width=\textwidth]{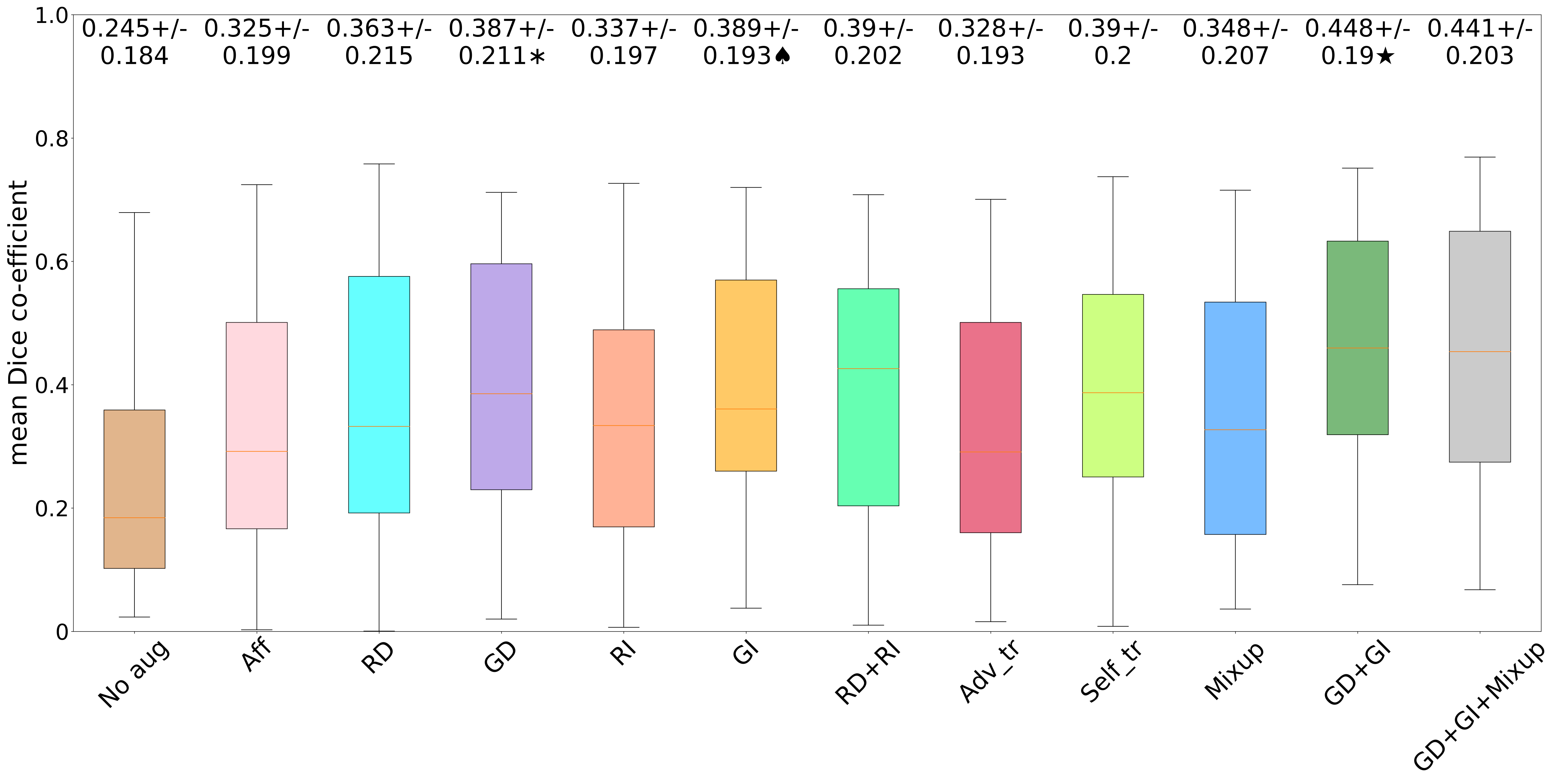}
        \caption{Peripheral Zone (Prostate data)}
        \label{fig:pz_res}
    \end{subfigure}
    \begin{subfigure}{0.9\textwidth}
        \includegraphics[width=\textwidth]{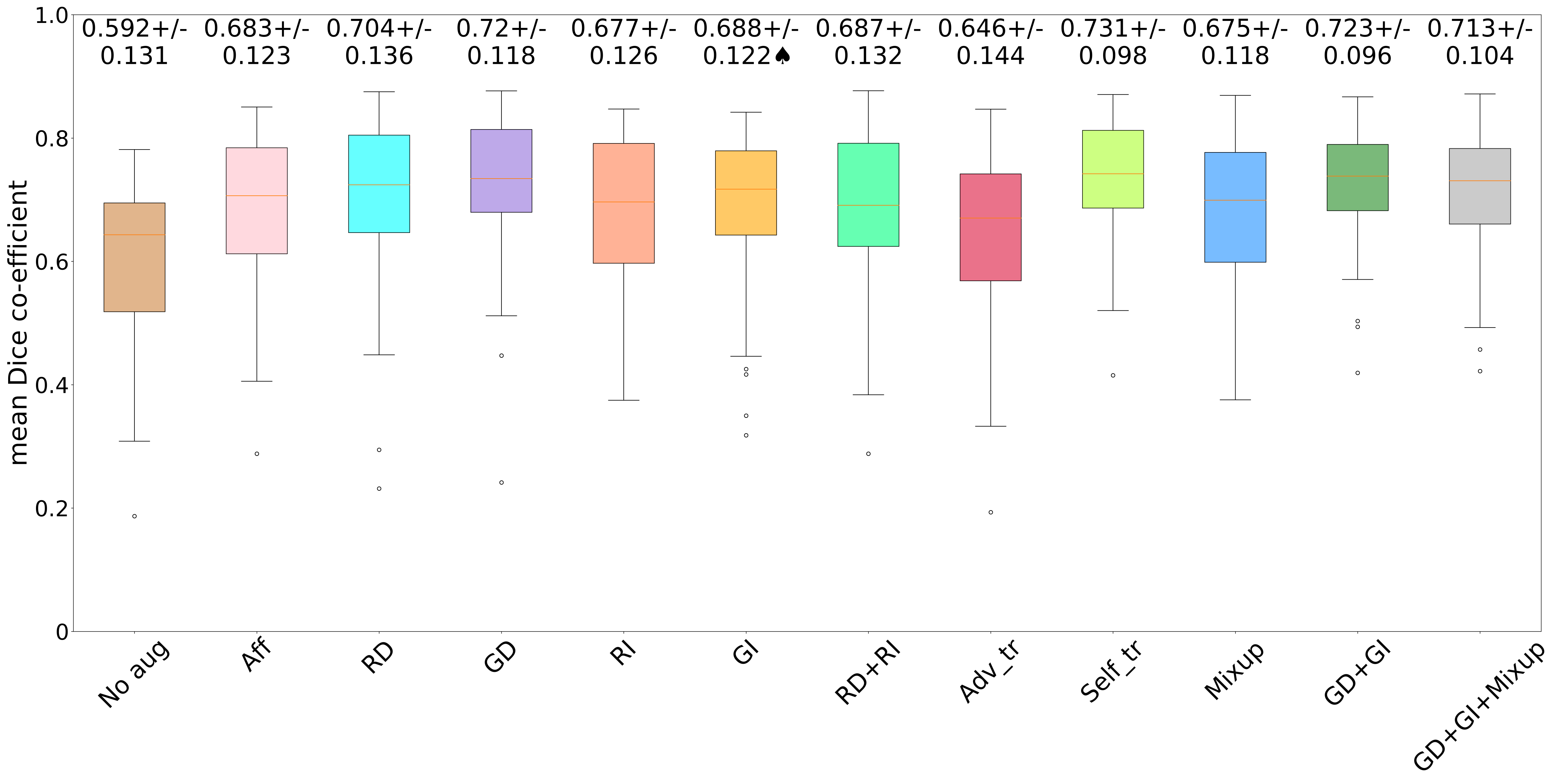}
        \caption{Central Gland (Prostate data)}
        \label{fig:cz_res}
    \end{subfigure}
    \begin{subfigure}{0.9\textwidth}
        \includegraphics[width=\textwidth]{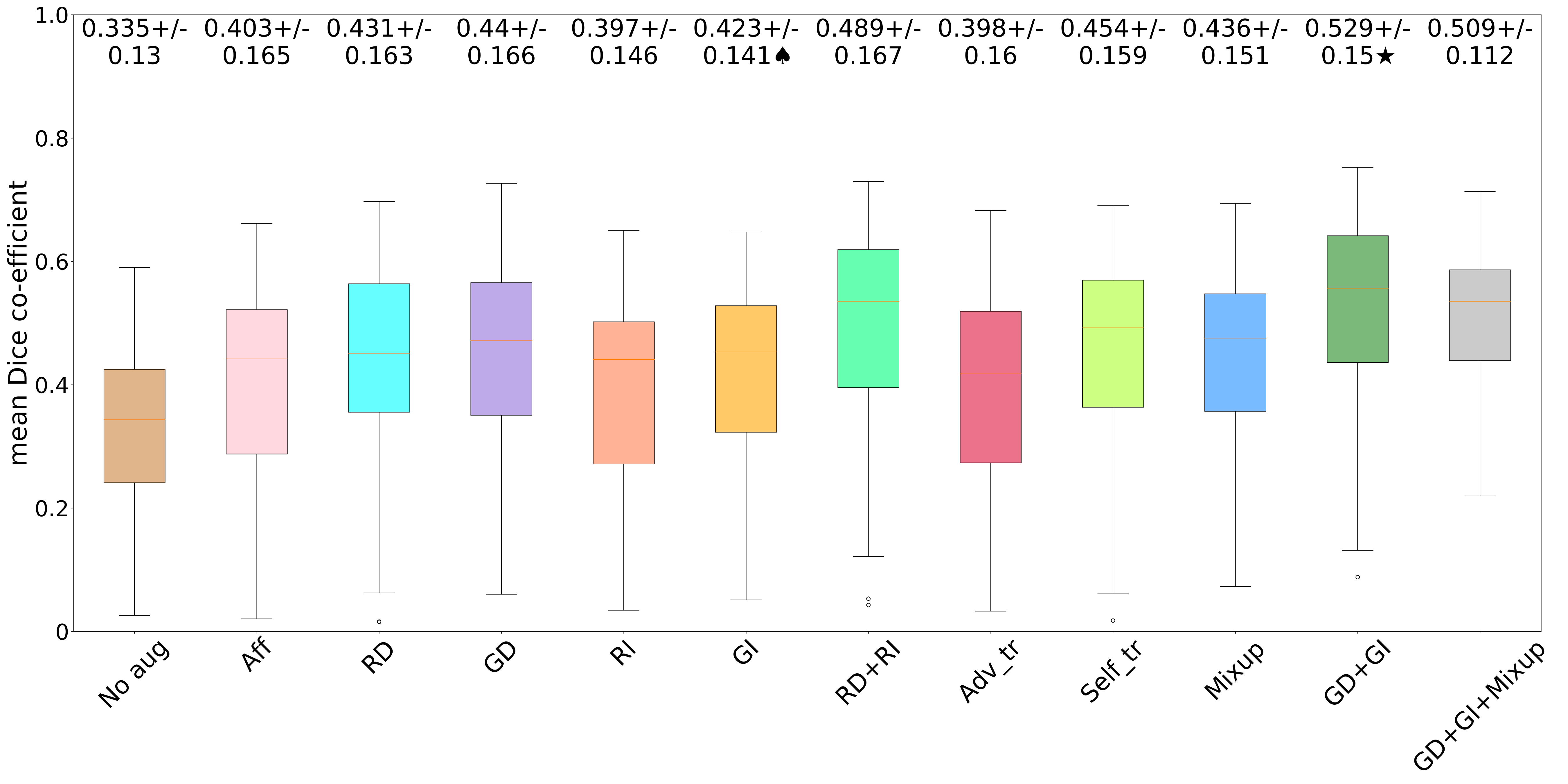}
        \caption{Pancreas+Tumor (Pancreas data)}
        \label{fig:panc_res}
    \end{subfigure}
    \caption{The segmentation performance of the proposed augmentation method (GD+GI) and several relevant works for three datasets are presented using Dice score (DSC). The number of labelled 3d volumes ($N_{L}$) used for training on cardiac, prostate, and pancreas datasets is one, one, and three, respectively (mean DSC and standard deviation values are reported on top of each boxplot). $\ast,\spadesuit,\star$ denotes the statistical significance of $GD$ over $RD$, $GI$ over $RI$, and $GD+GI$ over best performing related work, respectively (Wilcoxon signed rank test with threshold p value of 0.05).}
    %for $N_{L}=1$ for cardiac and prostate, and $N_{L}=3$ for pancreas datasets (mean DSC and standard deviation values are reported on top of each boxplot). $\ast,\spadesuit,\star$ denotes the statistical significance of $GD$ over $RD$, $GI$ over $RI$, and $GD+GI$ over best performing related work, respectively. (Wilcoxon signed rank test with threshold p value of 0.05) }
    %\vspace{-0.3cm}
    \label{fig:all_seg_res}
\end{figure}

\begin{figure}[!t]
    \centering
    \hspace{0cm}
    (a) \hspace{0.6cm}
    (b) \hspace{0.6cm}
    (c) \hspace{0.6cm}
    (d) \hspace{0.6cm}
    (e) \hspace{0.6cm}
    (f) \hspace{0.6cm}
    (g) \\
    \includegraphics[width=1.0\textwidth,trim={0cm 1cm 0cm 0.7cm},clip]{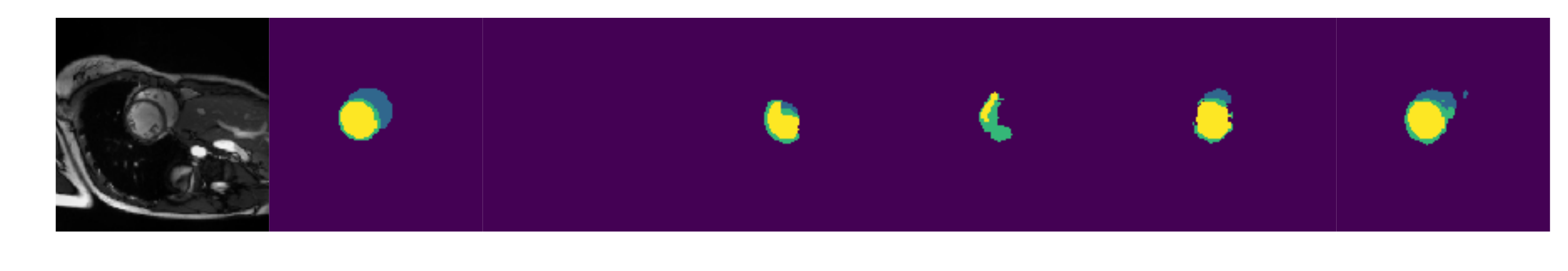}\\    
    \includegraphics[width=1.0\textwidth,trim={0cm 1cm 0cm 0.4cm},clip]{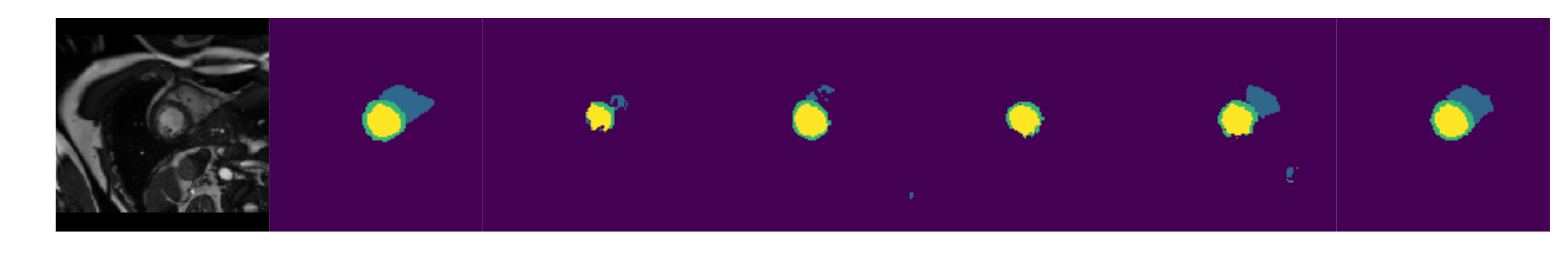}\\
    \includegraphics[width=1.0\textwidth,trim={0cm 1cm 0cm 0.4cm},clip]{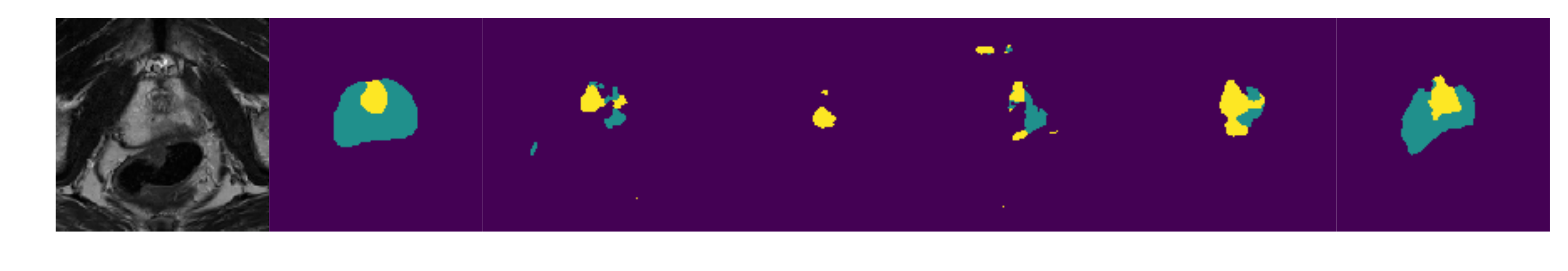}\\
    \includegraphics[width=1.0\textwidth,trim={0cm 1cm 0cm 0.4cm},clip]{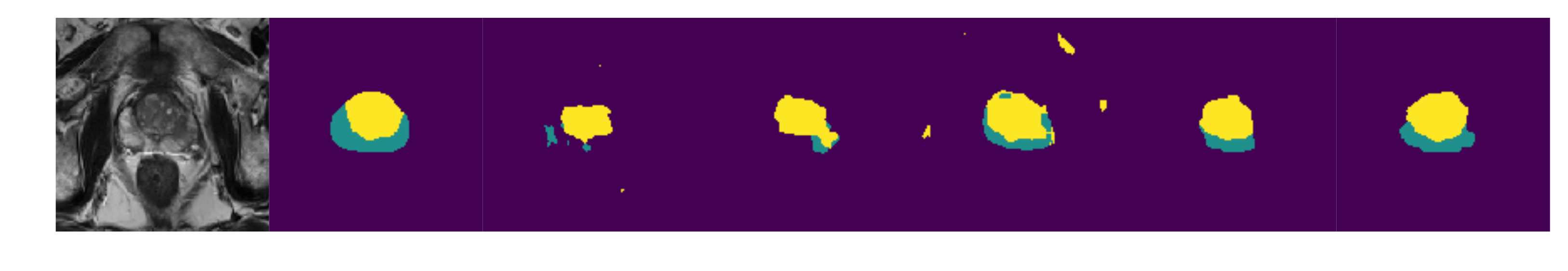}\\
    \includegraphics[width=1.0\textwidth,trim={0cm 1cm 0cm 0.4cm},clip]{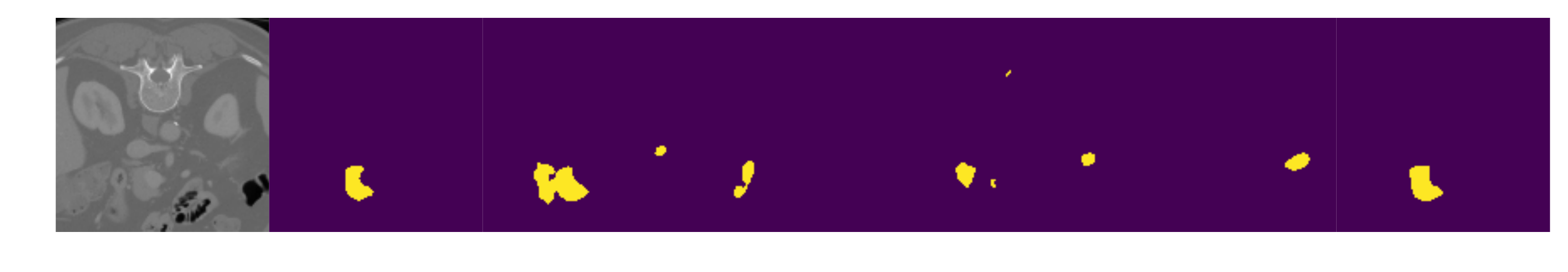}\\
    \includegraphics[width=1.0\textwidth,trim={0cm 1cm 0cm 0.4cm},clip]{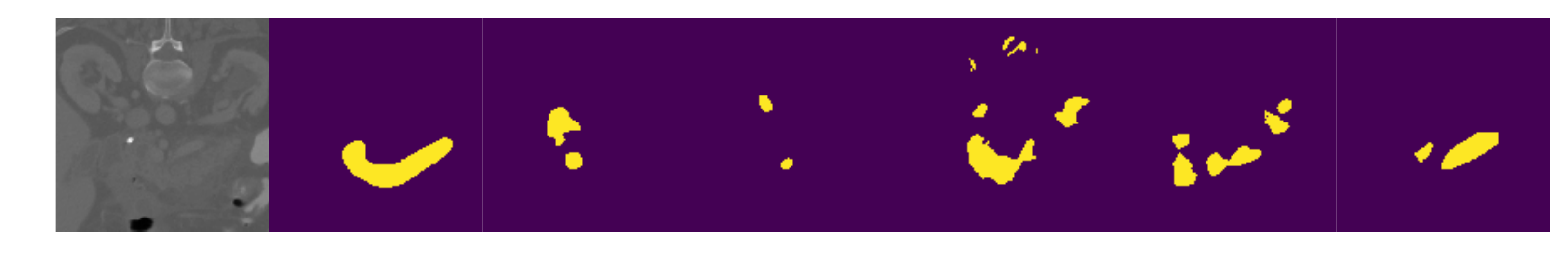}\\
    %\caption{Segmentation Results of proposed method vs earlier approaches}
    \caption{Qualitative comparison of the proposed method with other approaches is illustrated for two images each from cardiac, prostate, and pancreas datasets in the order of top to bottom: (a) input image, (b) ground truth, (c) Aff, (d) RD+RI, (e) Adv\_tr~\cite{zhang2017deep}, (f) Mixup~\cite{zhang2017mixup}, (g) GD+GI (Ours).}
    %\vspace{-0.3cm}
    \label{fig:seg_results}
\end{figure}

Figures (~\ref{fig:rv_res}-to-\ref{fig:lv_res}),(~\ref{fig:pz_res}-to-\ref{fig:cz_res}) and~\ref{fig:panc_res} present the quantitative results of the experiments on the cardiac, prostate and pancreas datasets, respectively.
The mean DSC and standard deviation values over all the runs are reported on the top of each boxplot in these figures.
We observe that the proposed method of augmentation (i.e.,GD+GI) outperforms the other data augmentation and semi-supervised learning methods considered here.
For qualitative inspection, we present some examples of visual results in Fig.~\ref{fig:seg_results}.
In the rest of this section, we present further analysis of the experimental results.

%Affine data augmentation
As expected, the lowest performance was observed when no data augmentation is used. Employing affine augmentation alone provided a substantial boost in performance. Adding random elastic deformations or random intensity fluctuations on top of affine augmentation yielded further improvements.% in accuracy.

\begin{figure}[!t]
    \centering
    \includegraphics[width=1.0\textwidth,trim={0cm 9.35cm 0cm 0.6cm},clip]{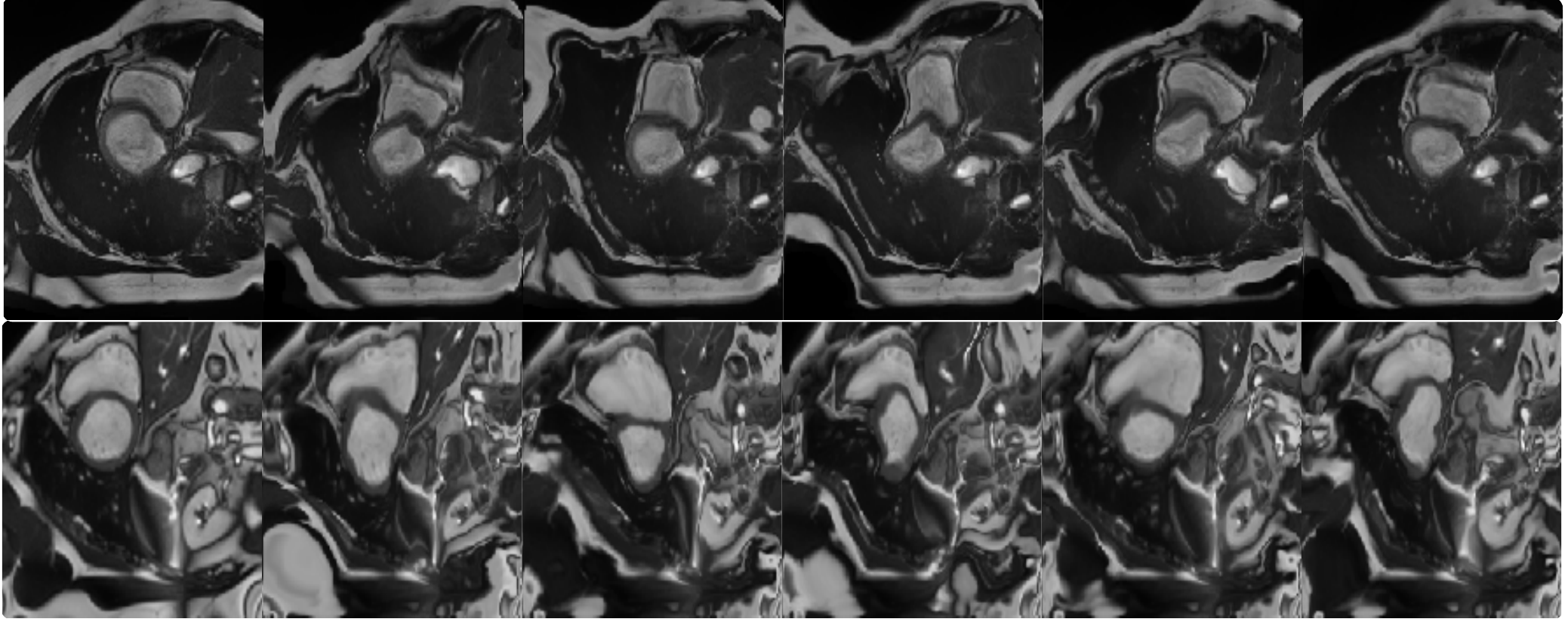}
    \\ \hspace{-2.65cm} Input Image \hspace{-0.15cm} $|$ Images generated by $G_V$ $\rightarrow$ \\
    \includegraphics[width=1.0\textwidth,trim={0cm 0cm 0cm 9.6cm},clip]{images/geo_trans_imgs/geo_int_trans_imgs.pdf}
    \\ \hspace{-2.65cm} Input Image \hspace{-0.15cm} $|$ Images generated by $G_I$ $\rightarrow$
    \caption{Generated augmentation images from the deformation field generator $G_V$ (top) and the intensity field generator $G_I$ (bottom) for the cardiac dataset.
    %Top image contains generated images with deformation fields and bottom image contains generated images with additive intensity fields applied
    }
    %\vspace{-0.3cm}
    \label{fig:gen_geogan_imgs}
\end{figure}

%DFGAN
Using the proposed learned deformation fields(GD) for augmentation yielded higher performance compared to random elastic deformations(RD). 
This results suggest that the proposed approach to learn a deformation field generator, by optimizing the segmentation accuracy along with a regularization term that leverages unlabeled examples, provided augmented examples more useful for obtaining high segmentation performance than random deformations.
Some samples of the generated deformed images are illustrated in the Fig.~\ref{fig:gen_geogan_imgs}. Surprisingly, we observed that the generated images did not always have realistic anatomical shapes. This is contrary to the popular belief, but generating realistic images may not be necessary nor optimal to obtain the best segmentation network.

%AIFGAN
Similar to the deformation case, the proposed additive intensity mask(GI) based augmentations performed better than random intensity fluctuations(RI). 
Here as well, the result suggest benefits of optimizing the intensity transformation generator using the proposed approach. 
Fig.~\ref{fig:gen_geogan_imgs} illustrates that the images generated from the learned intensity transformation generator are not necessarily realistic.

%Using both DFGAN and AIFGAN images
Since both the generators $G_V$ and $G_I$ are modeled to encapsulate different characteristics of the entire population, using both the augmentations is expected to produce higher performance gains over using only one of the augmentation separately. In our experiments, we indeed observed a substantial improvement in DSC when both are used together.
%Satisfying the expectation, we observed a substantial improvement in DSC when both are used together in our experiments.

%SSL methods perf
%Mixup
To our surprise, we observed that the Mixup augmentation yielded substantial performance gain over the affine transformations and random elastic deformations. This improvement was despite the augmented images being unrealistic. We attribute the performance improvement to the fact that Mixup creates soft probability maps for augmented images, which has been hypothesized to assist the optimization by providing additional information for training samples~\cite{hinton2015distilling}. 
%mixup + our methods
Applying Mixup over the augmented data obtained from the trained generators $G_V$ and $G_I$ yielded further marginal improvements, which suggests both approaches have complementary benefits.% for augmentation.

%self-training
With self-training, we observed an improvement in DSC over the affine augmentations as illustrated in Fig.~\ref{fig:all_seg_res}. This has been well-documented in SSL literature~\cite{yarowsky1995unsupervised,li2005setred} that re-training the neural network with the estimated predictions of the unlabeled data can assist in improving the segmentation performance~\cite{bai2017semi} in limited annotation setting. 
Although this yields some improvement over the affine augmentations, it did not outperform the proposed augmentations (GD+GI) except for the structure central gland of the prostate dataset. 
%This method still does not outperform the proposed augmentations (GD+GI).

%adversarial training - image-wise & pixel-wise
The semi-supervised adversarial training~\cite{souly2017semi,zhang2017deep} provided marginal performance gains over the baseline with affine augmentation (Results of ~\cite{souly2017semi} that uses image-level adversarial training are not reported as the GAN training did not converge to reasonable performance for the case of one labeled volume). This observation is not surprising as it has also been shown for other tasks in~\cite{oliver2018realistic}, SSL training may yield minimal performance gains when affine augmentations are included in the training.

\begin{figure}[!ht]
    \includegraphics[width=0.9\textwidth,trim={0.2cm 0cm 0.2cm 0.2cm},clip]{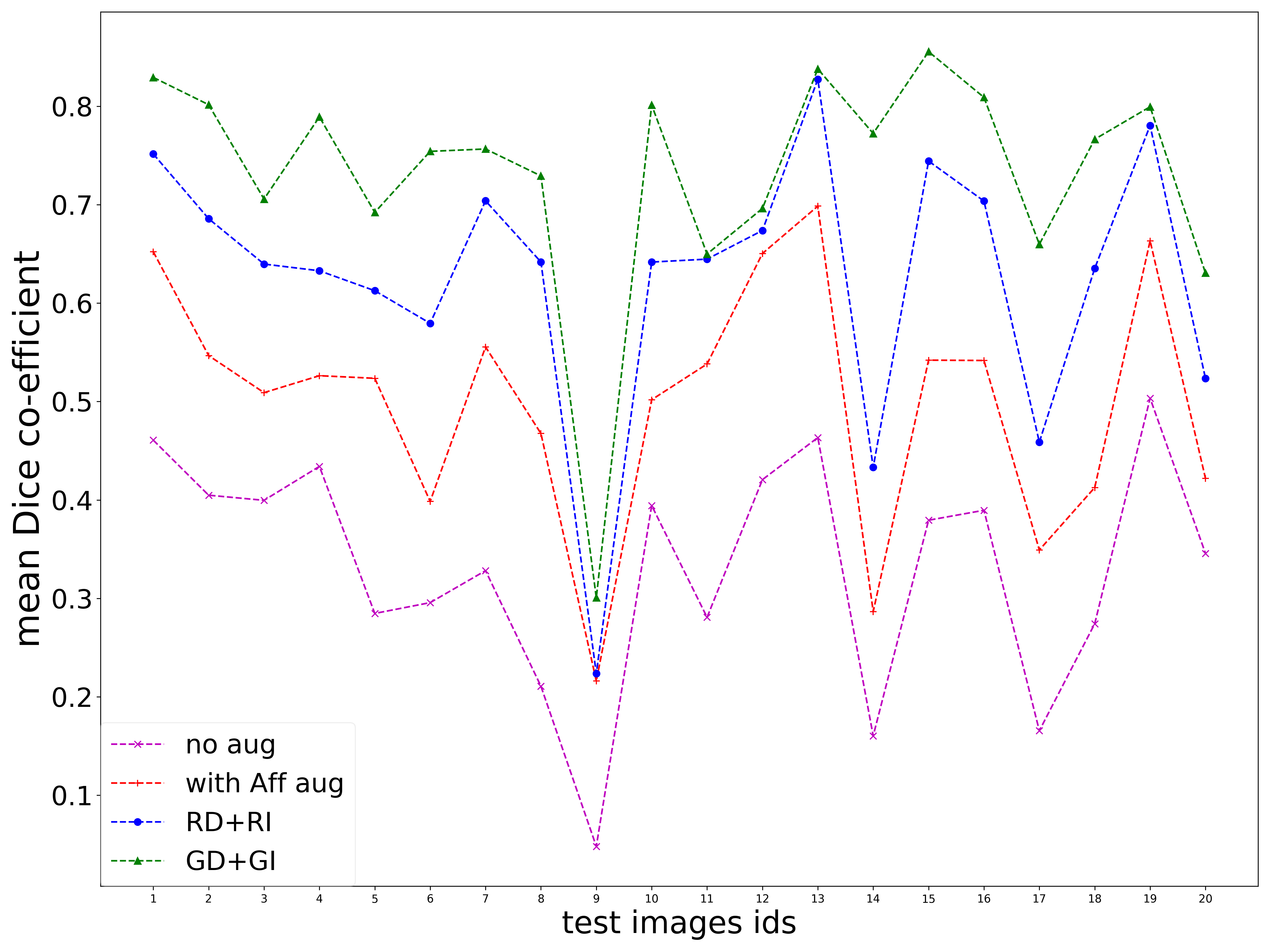}\\
    (a) Cardiac dataset (mean dice per test subject)\\
    \includegraphics[width=0.9\textwidth,trim={0.2cm 0cm 0.2cm 0.2cm},clip]{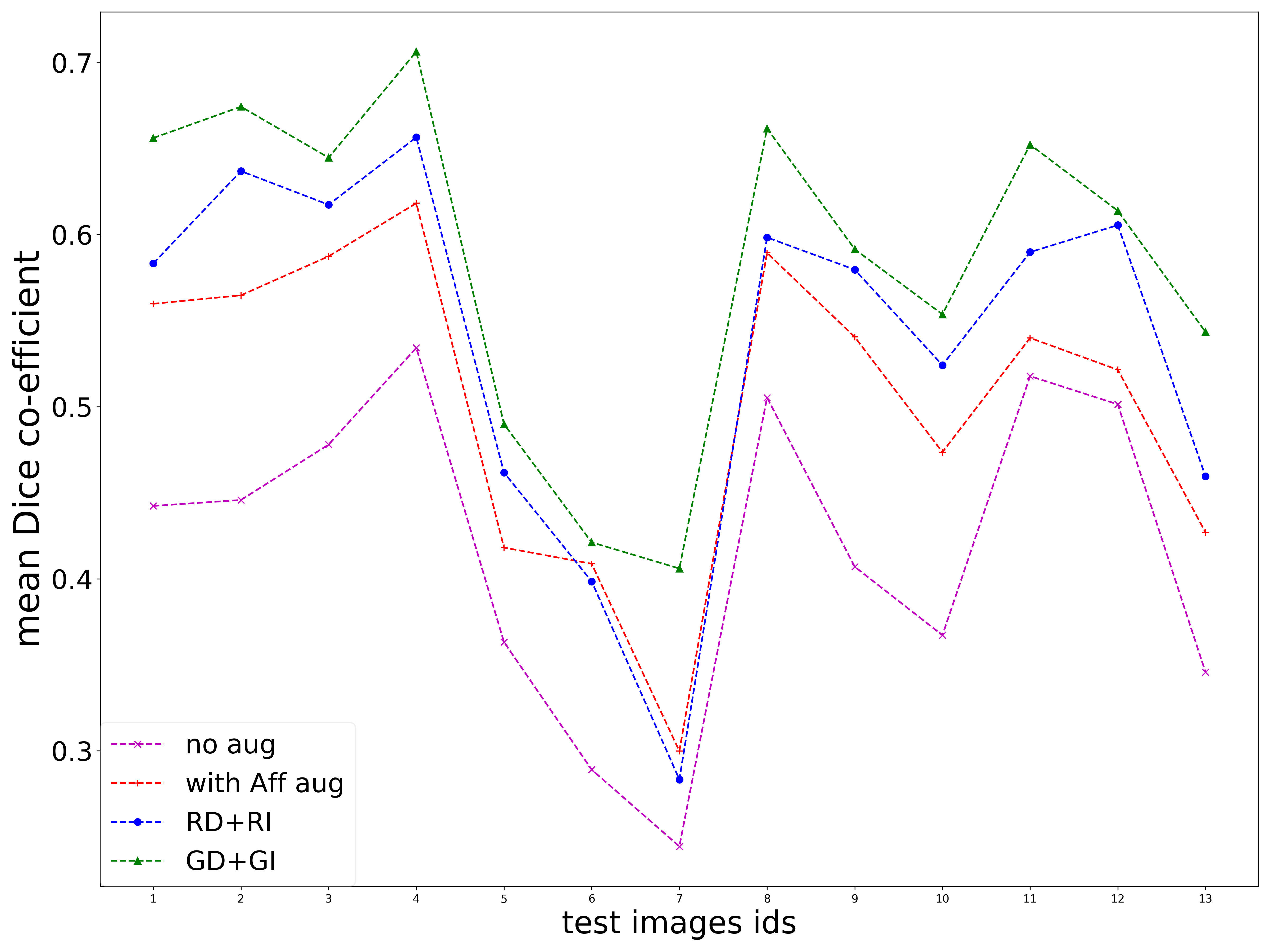}\\
    (b) Prostate dataset (mean dice per test subject)\\
    \includegraphics[width=0.9\textwidth,trim={0.2cm 0cm 0.2cm 0.2cm},clip]{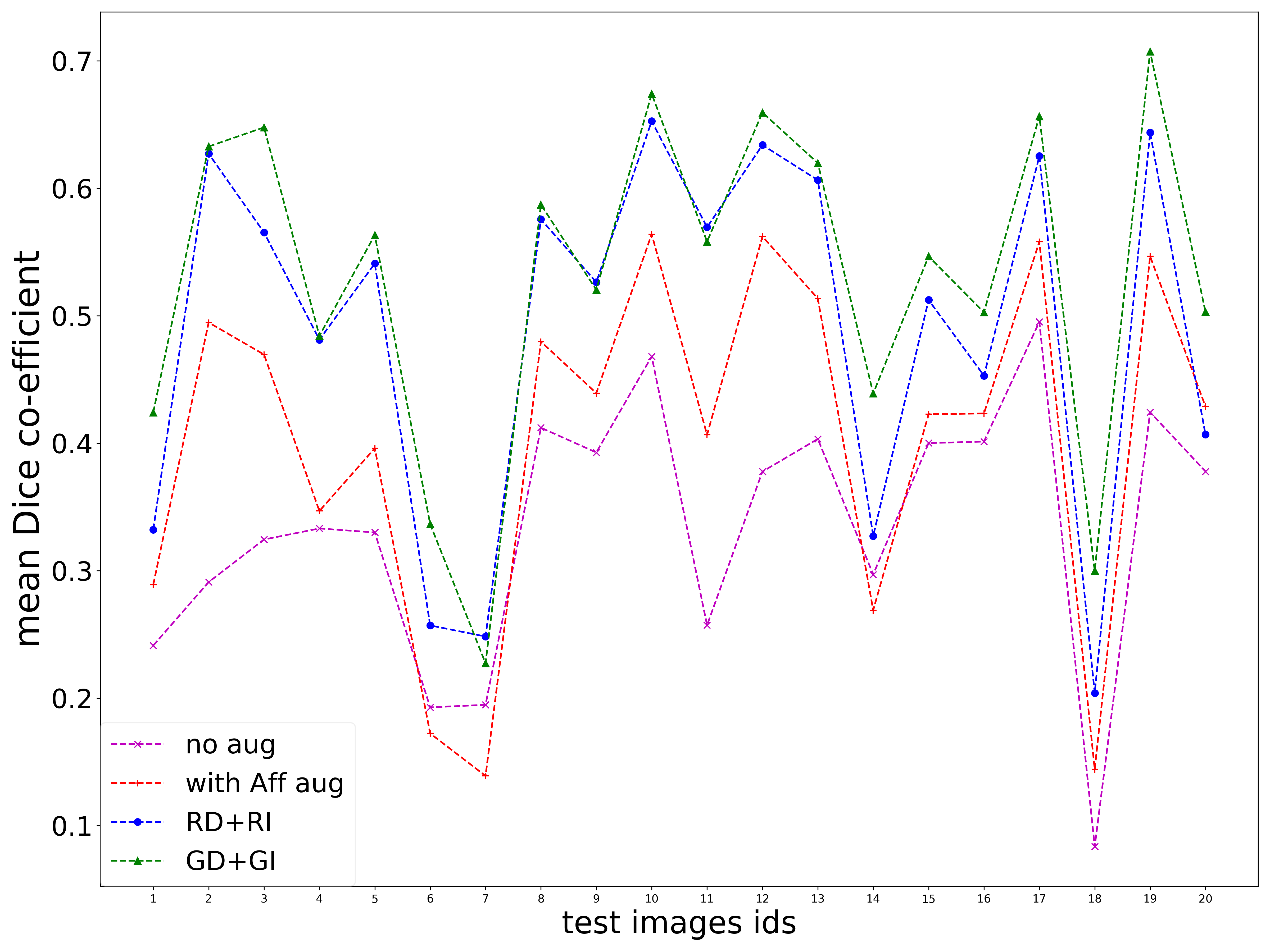}\\
    (c) Pancreas dataset (mean dice per test subject)\\
    \caption{Comparison of affine, random, and proposed augmentations' performance for each test subject. Mean Dice scores over 15 runs for each test subject is presented for $N_L=1$ for the cardiac,prostate dataset and $N_L=3$ for the pancreas dataset.}
    \vspace{-0.1cm}
    \label{fig:dsc_per_each_subj}
\end{figure}

\begin{figure}[!t]
    \centering
    \hspace{-0.1cm}
    (a) image ($X$) \hspace{0.55cm} 
    (b) deformation field \hspace{0.1cm} 
    (c) transformed \hspace{0.0cm} \\
    \hspace{2.7cm}
    over image ($\textbf{v}$) \hspace{1.2cm}
    image ($X \circ \textbf{v}$) \\
    \includegraphics[width=1.0\textwidth,trim={0cm 0.5cm 0cm 0.5cm},clip]{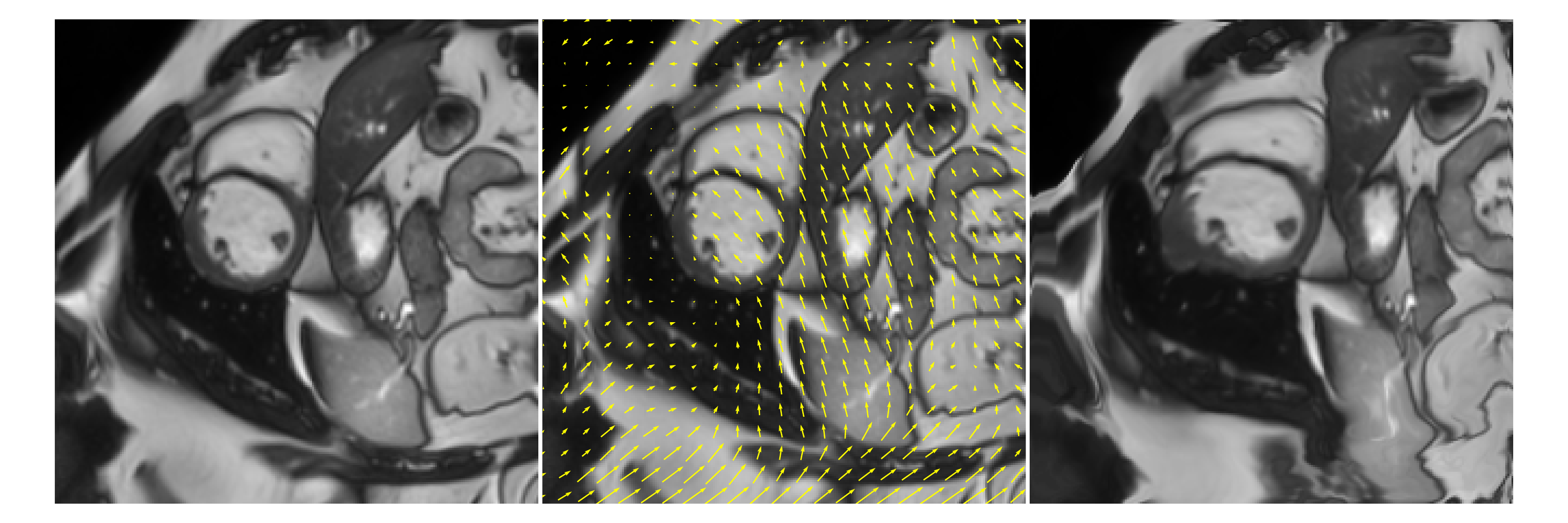}\\    
    \includegraphics[width=1.0\textwidth,trim={0cm 0.5cm 0cm 0.4cm},clip]{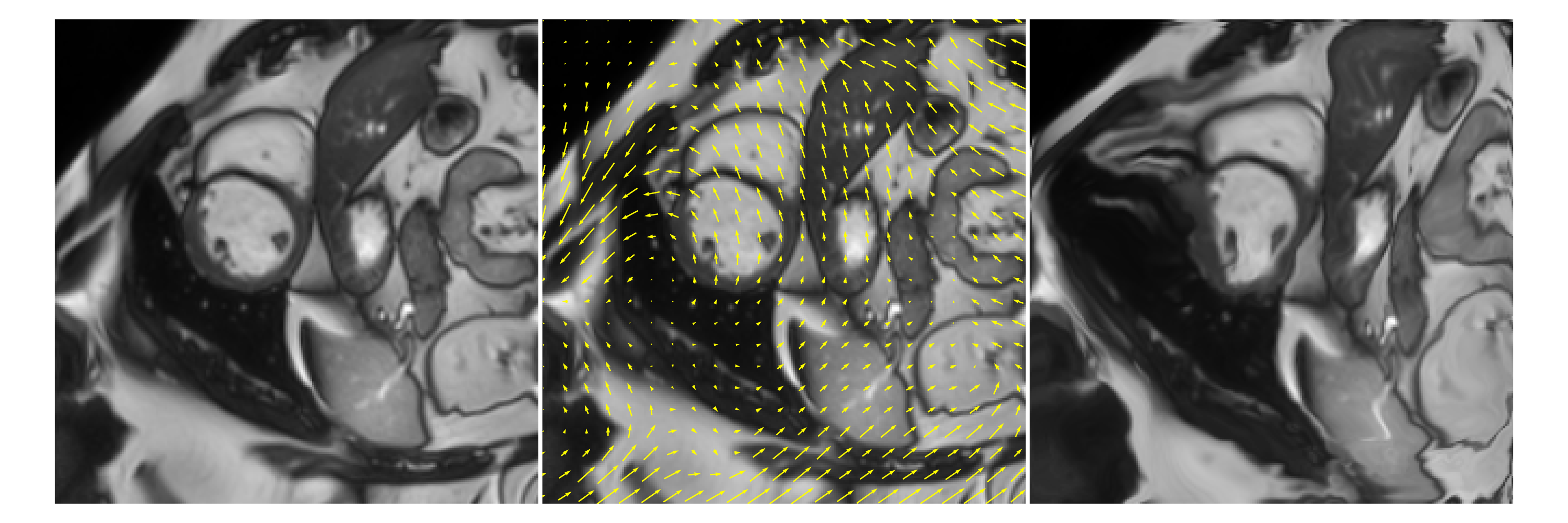}\\
    \includegraphics[width=1.0\textwidth,trim={0cm 0.5cm 0cm 0.4cm},clip]{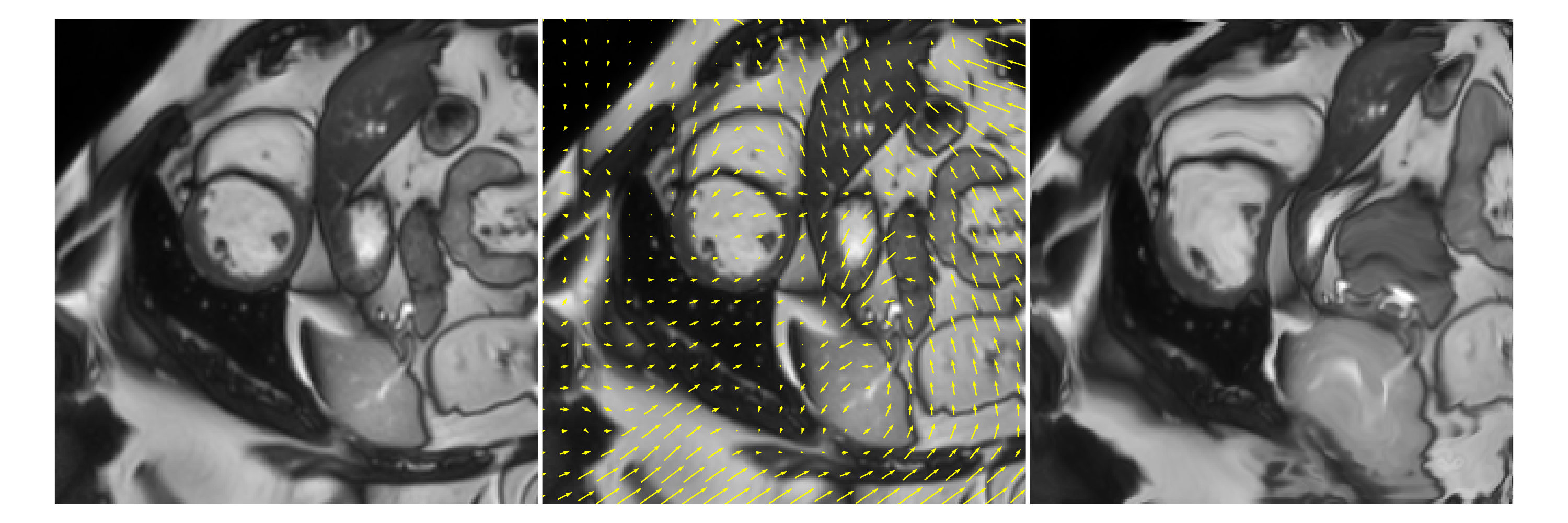}\\
    (i) Transformed images are obtained by applying deformation fields produced by Generator $G_V$.
    
    \includegraphics[width=1.0\textwidth,trim={0cm 0.5cm 0cm 0.5cm},clip]{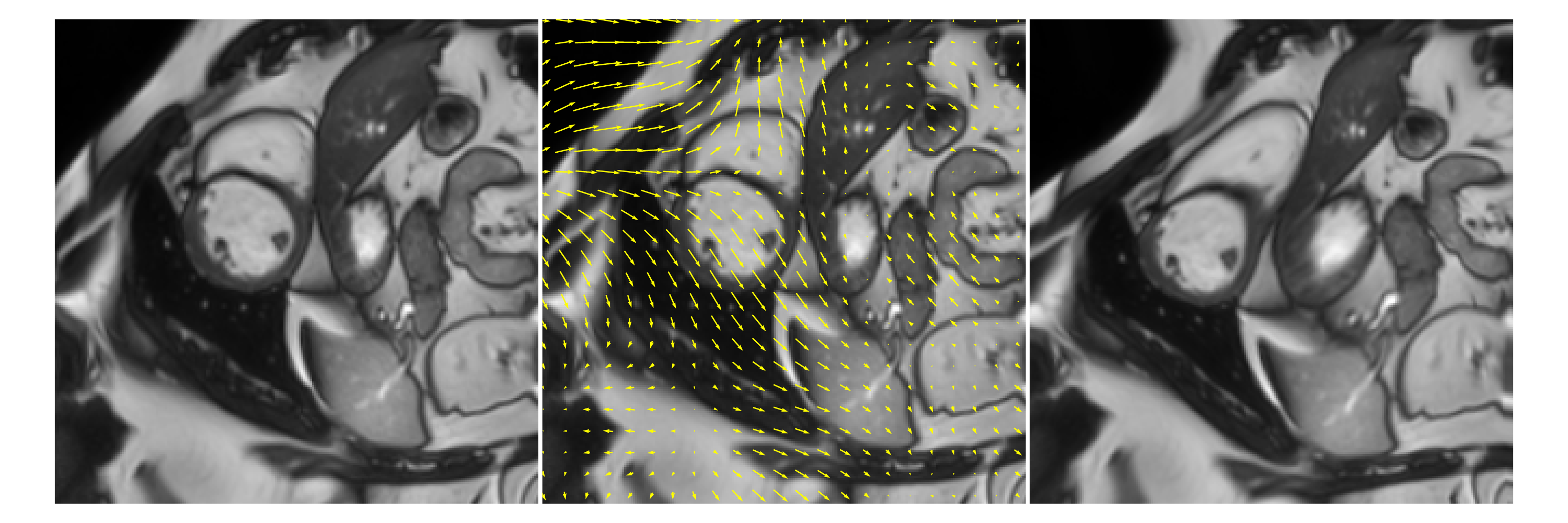}\\    
    \includegraphics[width=1.0\textwidth,trim={0cm 0.5cm 0cm 0.4cm},clip]{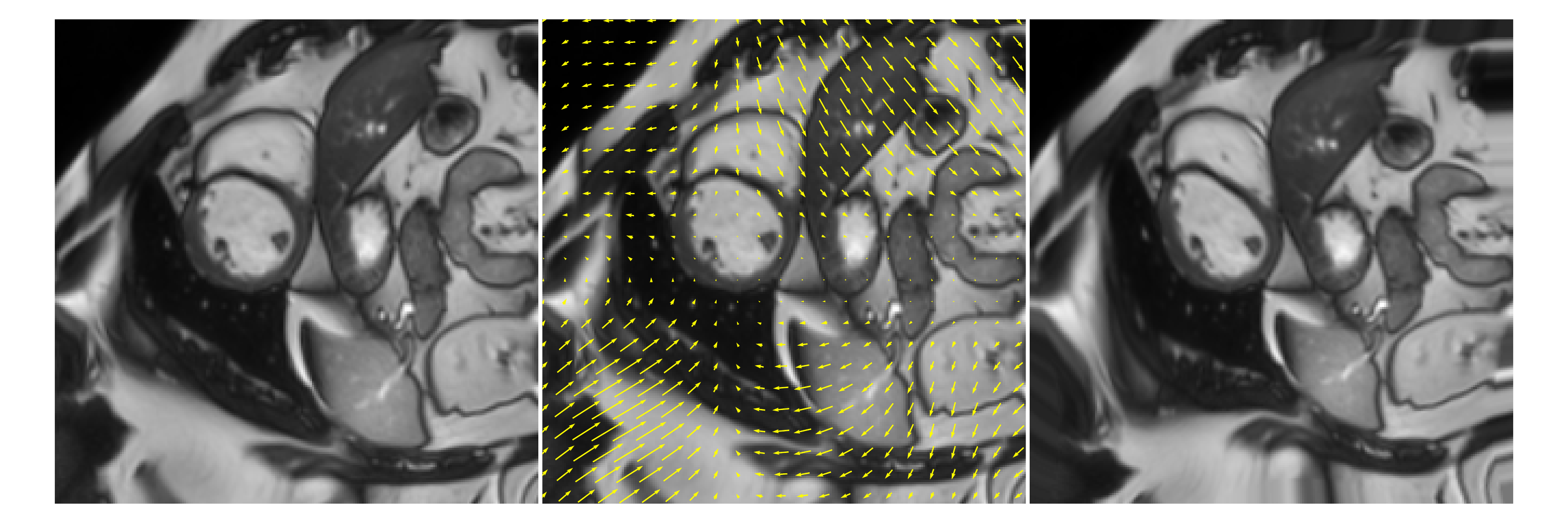}\\
    \includegraphics[width=1.0\textwidth,trim={0cm 0.5cm 0cm 0.4cm},clip]{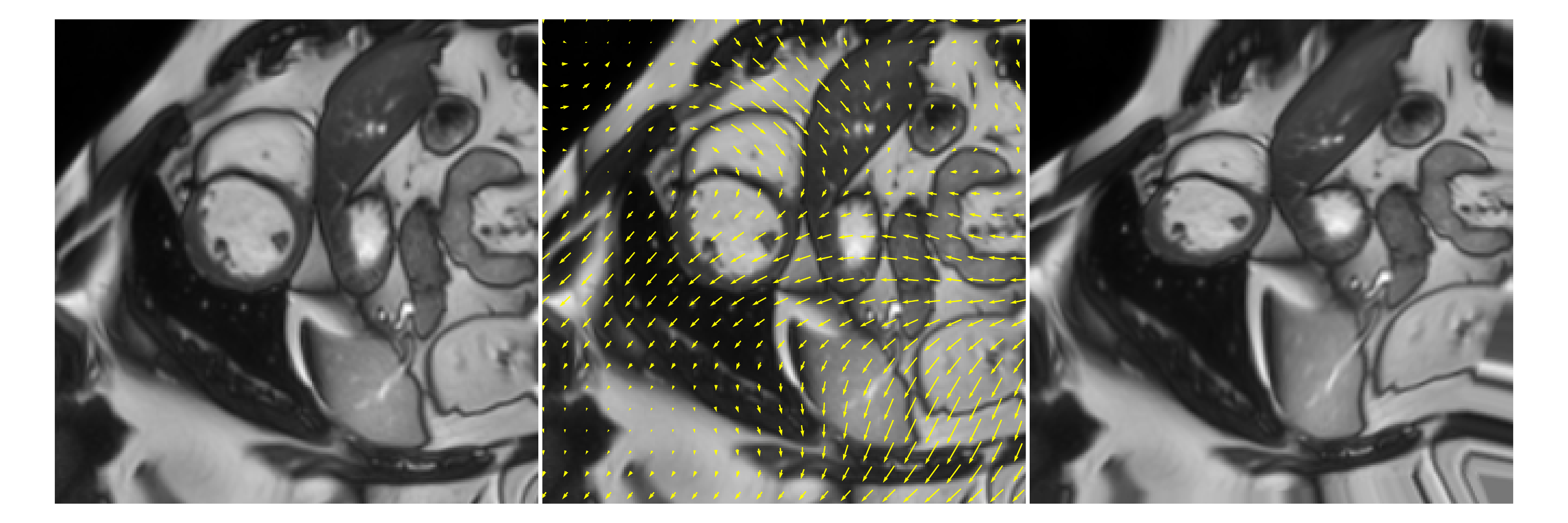}\\
    
    (ii) Transformed images are obtained by applying random elastic deformations.
    \caption{Generated deformation fields ($\textbf{v}$) and corresponding shape transformed augmentation images ($X \circ \textbf{v}$) obtained from the deformation field generator $G_V$ or random elastic deformations for an input image ($X$) from the cardiac dataset.}
    \vspace{-0.1cm}
    \label{fig:geo_trans_fields}
\end{figure}

\begin{figure}[!t]
    \centering
    \hspace{-0.1cm}
    (a) image ($X$) \hspace{0.55cm} 
    (b) additive intensity \hspace{0.1cm} 
    (c) transformed \hspace{0.0cm} \\
    \hspace{2.7cm}
    field ($\Delta I$) \hspace{1.2cm}
    image ($X + \Delta I$) \\
    \includegraphics[width=1.0\textwidth,trim={0cm 0.5cm 0cm 0.5cm},clip]{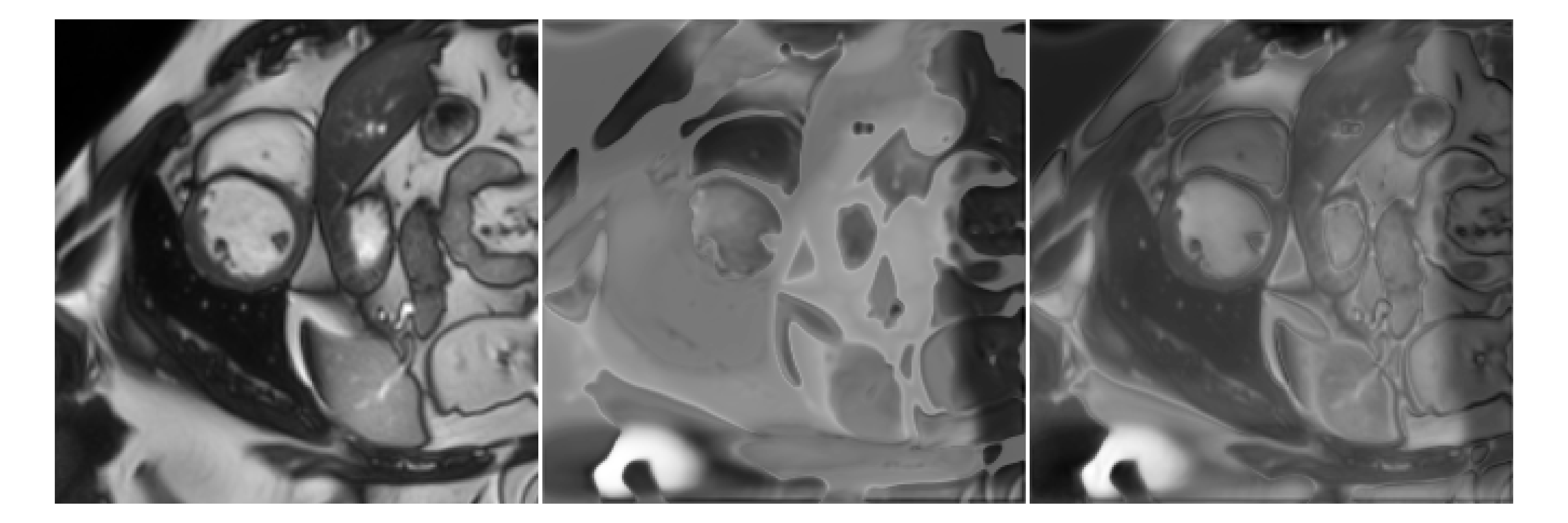}\\    
    \includegraphics[width=1.0\textwidth,trim={0cm 0.5cm 0cm 0.4cm},clip]{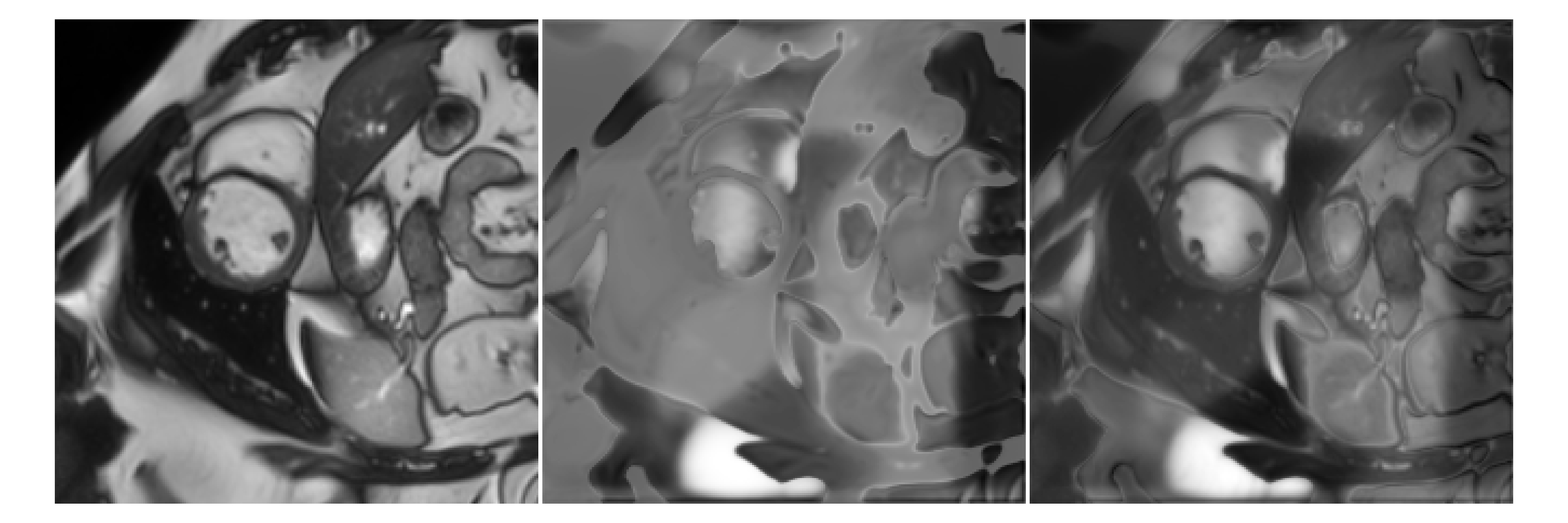}\\
    \includegraphics[width=1.0\textwidth,trim={0cm 0.5cm 0cm 0.4cm},clip]{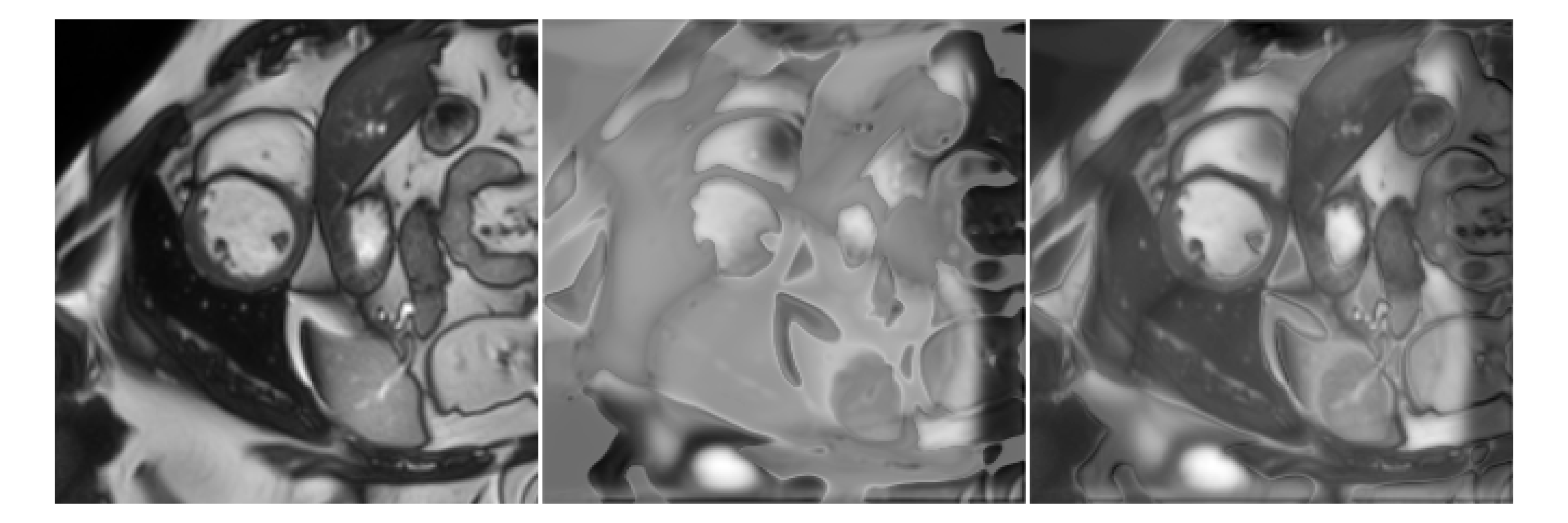}\\
    (i) Transformed images are obtained by applying additive intensity fields produced by Generator $G_I$. 
    \includegraphics[width=1.0\textwidth,trim={0cm 0.5cm 0cm 0.5cm},clip]{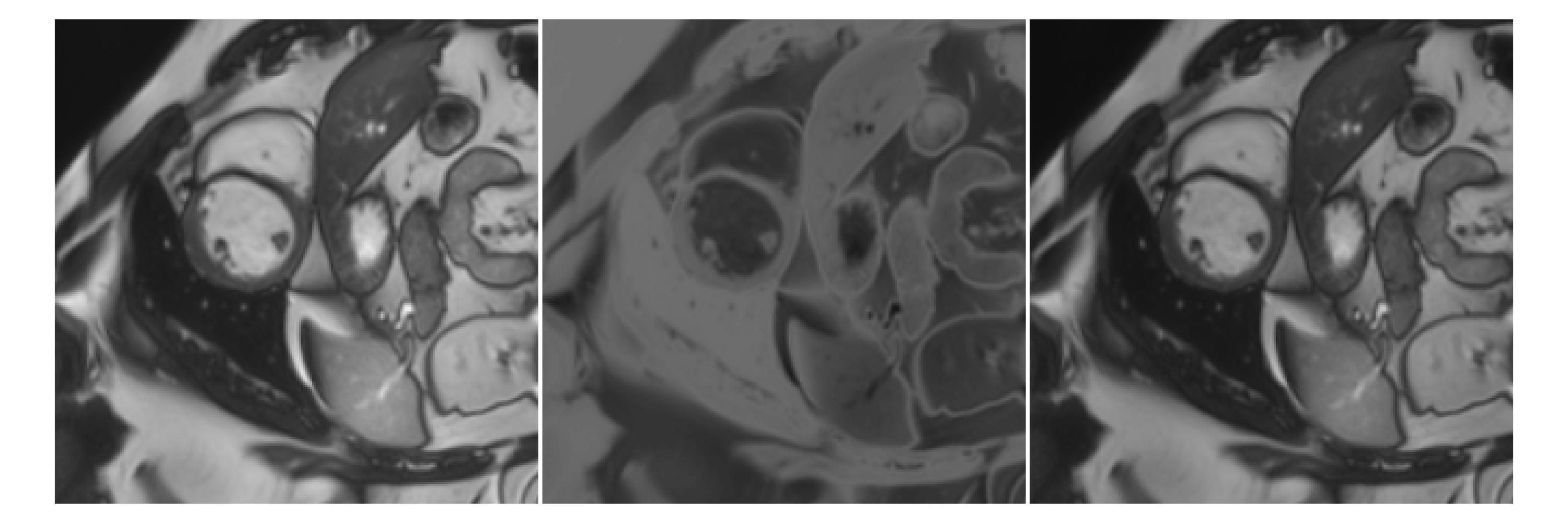}\\    
    \includegraphics[width=1.0\textwidth,trim={0cm 0.5cm 0cm 0.4cm},clip]{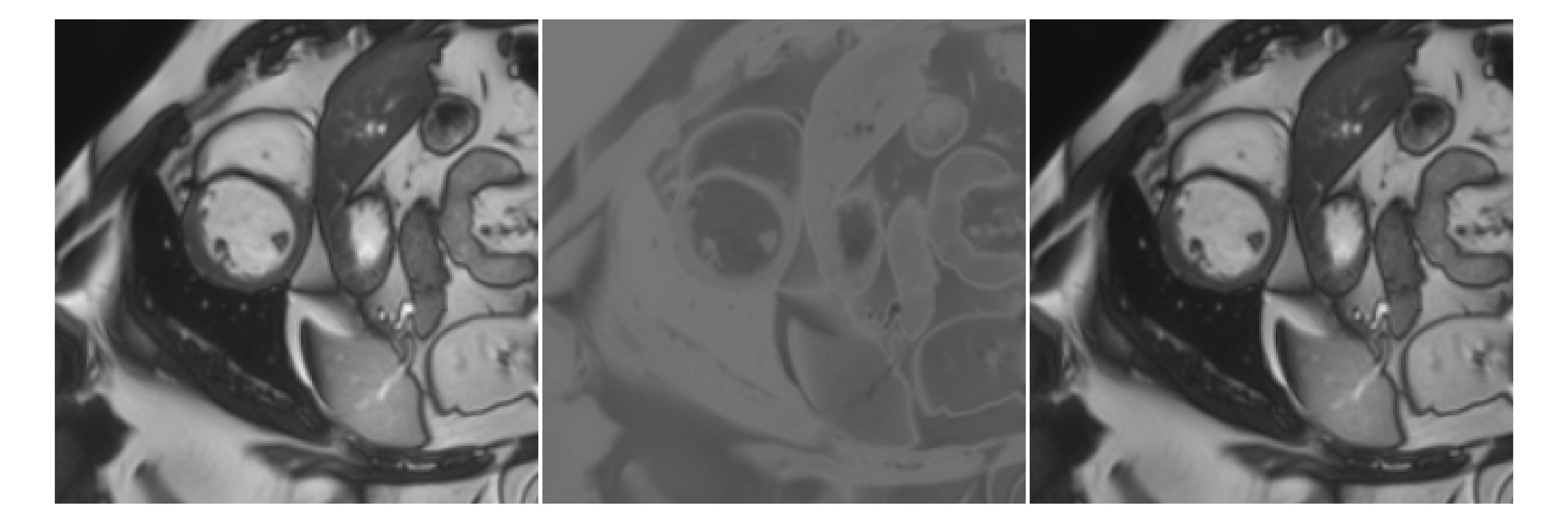}\\
    \includegraphics[width=1.0\textwidth,trim={0cm 0.5cm 0cm 0.4cm},clip]{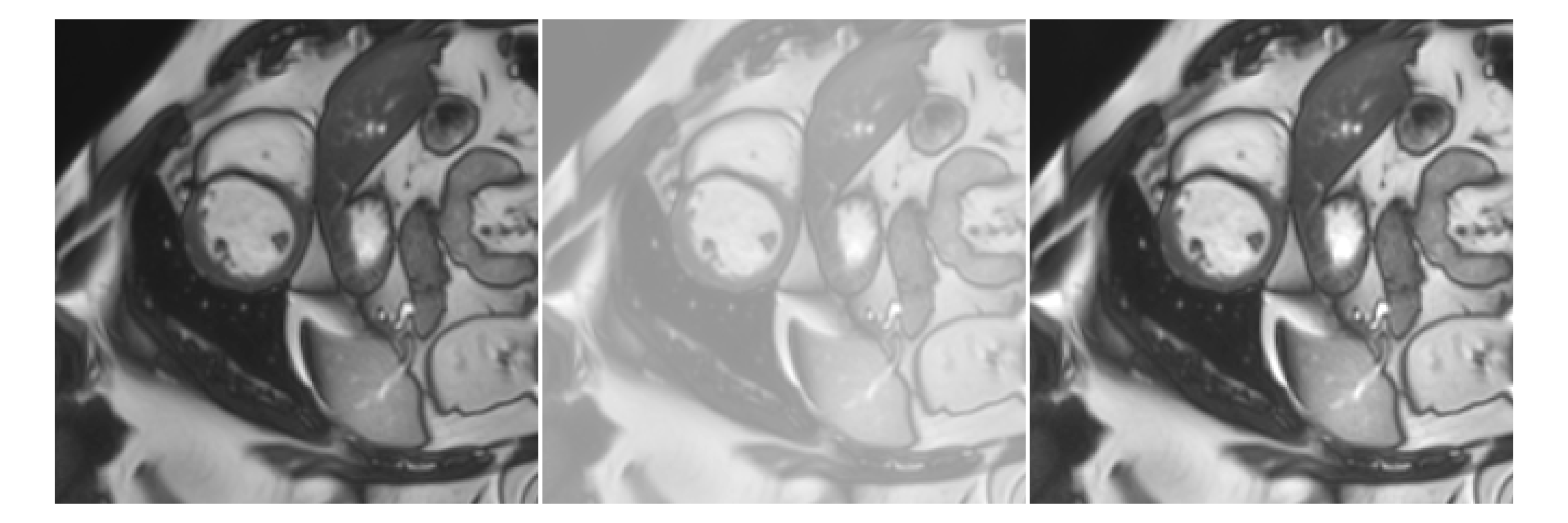}\\
    (ii) Transformed images are obtained by applying random contrast and brightness values.
    \caption{Generated additive intensity fields ($\Delta I$) and corresponding intensity transformed augmentation images ($X + \Delta I$) obtained from the intensity field generator $G_I$ or random contrast and brightness for an input image ($X$) from the cardiac dataset.}
    \vspace{-0.1cm}
    \label{fig:int_trans_fields}
\end{figure}

In the Appendix, we provide additional analysis plots such as: (a) the performance improvement seen per test subject averaged over all the runs for each dataset in Fig.~\ref{fig:dsc_per_each_subj}. We observed that for the majority of test subjects, the proposed method performs better or equal to random augmentations.
(b) A sample of generated deformation and additive intensity fields obtained from the trained generators on the cardiac dataset are provided in Fig.~\ref{fig:geo_trans_fields} and Fig.~\ref{fig:int_trans_fields}, respectively.

Despite getting performance improvements using the proposed method for the three datasets evaluated in this work, it is important to note that it is a computationally expensive method to deploy on any new dataset. This is because one would need to search for the optimal hyper-parameters ($\lambda_{adv}$, $\lambda_{LD}$) for the loss terms.

Also, we want to state that we do not think our model and experiments presented can provide any insight or address the intensity differences problem arising from scanner differences and nor does this study focuses on addressing this. 
We are modeling augmentations to maximize performance given a limited labeled and unlabeled set and evaluate the model with images from one type of scanner. Hence, we do not expect this setup to address the scanner differences problems.

\begin{figure}[!t]
    \centering
    \begin{subfigure}{1.0\textwidth}
        \includegraphics[width=8cm,height=4cm,trim={0.2cm 0cm 0.2cm 0.2cm}]{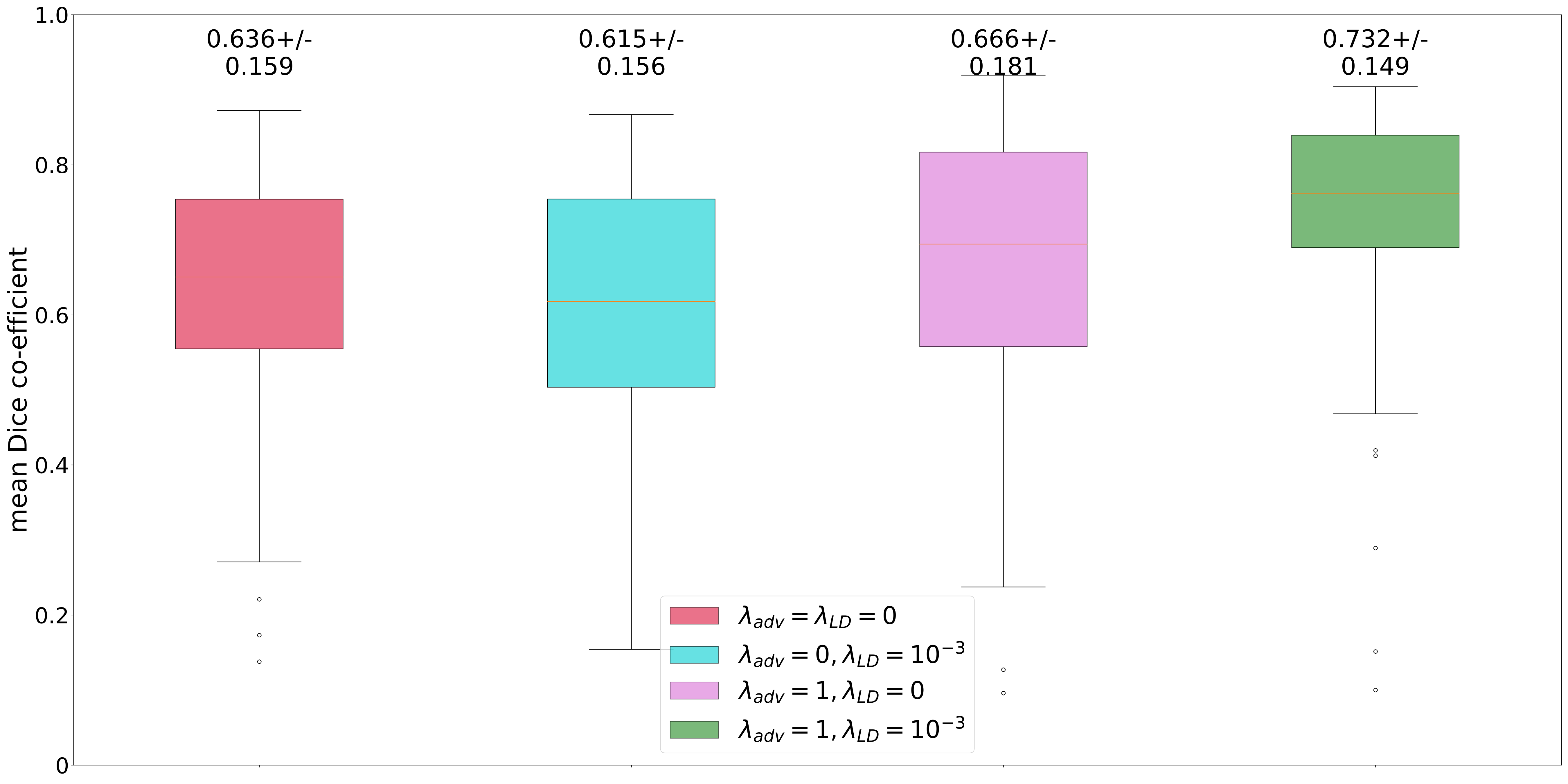}
        \caption{Effect of adversarial ($\lambda_{adv}$) and large deviation ($\lambda_{LD}$) loss terms of the regularization loss on the segmentation performance of the proposed method. The DSC with random augmentations (RD+RI) is 0.627 from Fig. 1 for reference to compare with the combinations of loss terms evaluated here.}
        \label{fig:vary_terms}
    \end{subfigure}
    \begin{subfigure}{1.0\textwidth}
        \includegraphics[width=8cm,height=4cm,trim={0.2cm 0cm 0.2cm 0.2cm}]{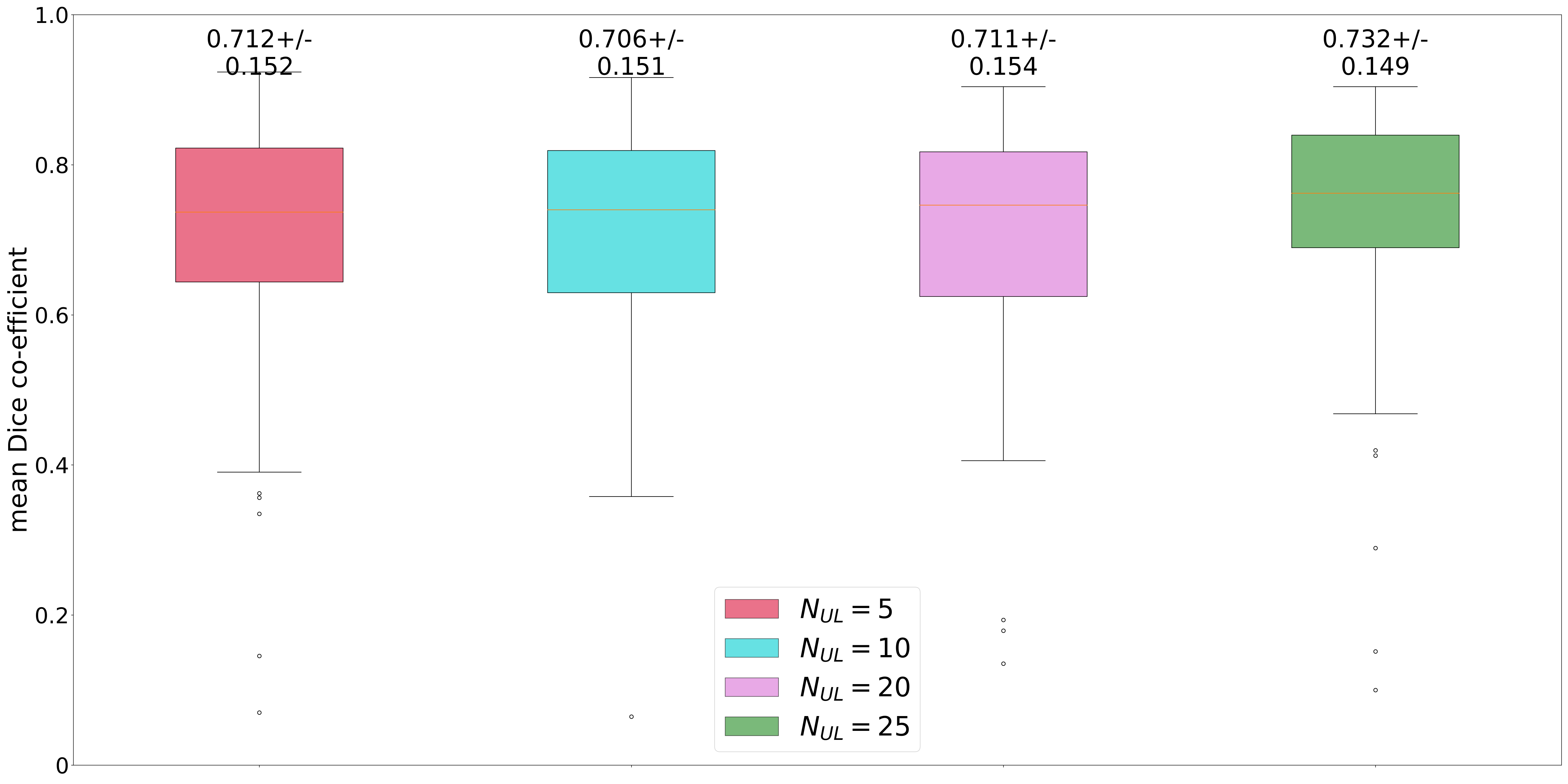}
        \caption{Effect of number of 3D unlabeled images ($N_{UL}$) on the segmentation performance}
        \label{fig:vary_unlabeled_data}
    \end{subfigure}
    \begin{subfigure}{1.0\textwidth}
        \includegraphics[width=8cm,height=4cm,trim={0.2cm 0cm 0.2cm 0.2cm}]{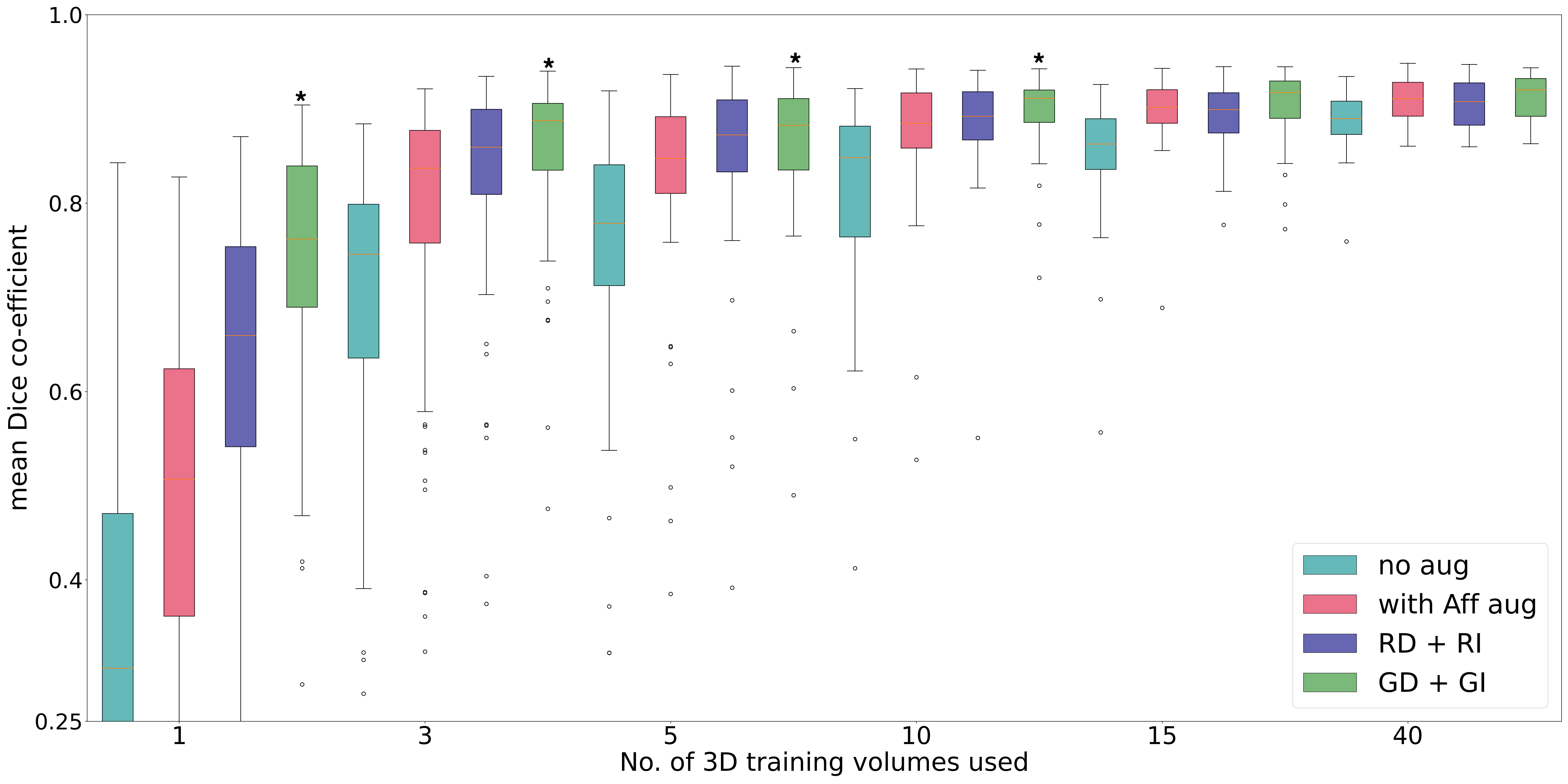}
        \caption{Effect of number of 3D labeled images ($N_L$) on the segmentation performance}
        \label{fig:vary_labeled_data}
    \end{subfigure}
    \caption{Results of segmentation performance on cardiac dataset quantifying: (a) effects of adversarial ($\lambda_{adv}$) and large deviation ($\lambda_{LD}$) loss terms of the regularization loss for $N_{L}=1$ (see~\ref{eff_terms_reg}-A), (b) effect of varying the number of 3D unlabeled images $N_{UL}$ for $N_{L}=1$ (see~\ref{vary_no_unl}-C) (mean DSC and standard deviation values reported on top of each boxplot) and (c) effect of varying the number of 3D labeled volumes ($N_{L}=1,3,5,10,15,40$) in the training (see~\ref{vary_no_lab}-D) and $\star$ here indicates statistically significant improvements between proposed method (GD+GI) against best of RD+RI and Aff augmentations using wilcoxon’s paired t-test with threshold value of 0.05.}
    %\vspace{-0.3cm}
    \label{fig:ablation_study_res}
\end{figure}

%\vspace{-0.1cm}
%Different terms of loss terms
%A. \textbf{Ablation study: effects of the terms of regularization loss}\label{eff_terms_reg}: 
\textbf{Ablation studies:}\\
\textbf{A. Effects of adversarial ($\lambda_{adv}$) and large deviation ($\lambda_{LD}$) loss terms of regularization loss on segmentation performance}\label{eff_terms_reg}:  In this experiment, we analyze the effect of each term of the regularization loss on the performance by varying the values of $\lambda_{adv}$ and $\lambda_{LD}$ in the training of the generators $G_V$ and $G_I$. 
%both adv loss & l1 loss - 0
Fig.~\ref{fig:vary_terms} presents the quantitative results of the analysis on the cardiac dataset for GD+GI augmentations. We observed the least performance gain over baseline when we disabled the whole regularization with $\lambda_{adv} = \lambda_{LD} = 0$, the setup similar to the work in~\cite{wang2018low}. This setup yielded performance similar to the case when both random deformations and intensity fluctuations were leveraged for augmentation.
%adv loss = 1 & l1 loss - 0
The performance gain we observe when we enable only the adversarial loss in the regularization, i.e. ($\lambda_{adv}=1,\lambda_{LD}=0$), can be attributed to enforcing the model to match the distribution of generated images to that of unlabeled images. This matching propels the generator to synthesize examples showing the diverse set of shape and intensity variations present in the unlabeled data.
%adv loss = 0 & l1 loss - 10e-3
We observed a deterioration in performance when only large-deviation loss is enabled on deformation and intensity fields ($\lambda_{adv}=0,\lambda_{LD}=10^{-3}$). This setup encourages the generators to explicitly produce larger deformation and intensity fields without any control from the adversarial term. Transformations output from these generators yields very unrealistic samples, in the most extreme case moving all the foreground pixels out of the frame. Such synthetic examples, naturally, are not useful for the segmentation task. 

%adv loss = 1 & l1 loss - 10e-3
We observed the highest performance boost when both terms were enabled($\lambda_{adv}=1,\lambda_{LD}=10^{-3}$). 
Effectively, for the proposed method to yield stated gains, one needs to use all three losses simultaneously, where they compete against each other. Also, this indicates the value of the combination of the terms and their complementary behavior.
The large-deviation loss, when used in addition to the adversarial loss, prevents the network from simply replicating the training data with generating relatively smaller transformations. Instead, it compels the generator to produce larger fields as long as they satisfy discriminator's objective. Effectively, producing augmented image-label pairs that are very different from the labeled image-label pairs. Adversarial loss on the other hand, contains the effects of the large-deviation loss by not allowing models to generate extreme transformations. 

%trim=left bottom right top
% \begin{figure}[t!]
%     \includegraphics[width=5cm,height=5cm,keepaspectratio,trim={0.2cm 0cm 0.2cm 0.2cm},clip]{images/seg_results/acdc/indiviual_loss_terms_opti_tr1_dsc_bar_plot.png}\\
%     \caption{Effect of the terms of the regularization loss on the segmentation performance(see~\ref{eff_terms_reg}), presented for the cardiac dataset (mean dice score and standard deviation values for $N_{L}=$1 on top of each barplot).}
%     \vspace{-0.1cm}
%     \label{fig:vary_terms}
% \end{figure}
%independent optimization
%\subsection{Independent optimization of the generator and the segmentation networks}\label{disjoint_opti}
\textbf{B. Independent optimization of the generator and the segmentation networks}\label{disjoint_opti}: One of the biggest claims of the proposed method is the benefit of using the segmentation cost function for learning the generators. 
This leads to a joint optimization of the segmentation network's parameters along with the generator's. 
Here, we examine the impact on segmentation performance when the generators are optimized without the segmentation loss, only using the regularization term with adversarial and large-deviation terms. 
Results shown in Table~\ref{tab:non_joint_opti} show that when the segmentation loss is included in the learning of the generators, i.e., joint optimization leads to much higher DSC values than independent optimization. 
Thus, empirically proving our point that task-driven optimization is better than the traditional independent optimization, which was the approach taken in earlier works~\cite{shin2018medical,bowles2018gan} to generate the augmented data independent of the down-stream task.
%\setlength{\belowcaptionskip}{-15pt}
%\setlength\belowcaptionskip{-1.5cm}
%\setlength\abovecaptionskip{-1.5cm}
% independent & 0.527 & 0.553 & 0.719 & 0.793 & 0.797 & 0.909 \\ 
%               & +/-0.266 & +/-0.218 & +/-0.23 & +/-0.187 & +/-0.097 & +/-0.09 \\ \hline
%     joint     & 0.651 & 0.710 & 0.834 & 0.832 & 0.823 & 0.922 \\ 
%               & +/-0.23 & +/-0.157 & +/-0.171 & +/-0.148 & +/-0.076 & +/-0.072 \\ \hline
\begin{table}[!b]
    \centering
    \begin{tabular}{|m{1.6cm}|m{0.77cm}|m{0.77cm}|m{0.77cm}|m{0.77cm}|m{0.77cm}|m{0.77cm}|}
    \hline
    %methods of& \multicolumn{6}{c|}{Number of 3D training volumes used}
    %\\ \cline{2-7} 
    %optimization & \multicolumn{3}{c|}{1} & \multicolumn{3}{c|}{3}
    %\\ \cline{2-7}
    methods of& \multicolumn{3}{c|}{$N_L=1$} & \multicolumn{3}{c|}{$N_L=3$}
    \\ \cline{2-7}
    optimization & RV & Myo & LV & RV & Myo & LV \\ \hline
    independent & 0.527 & 0.553 & 0.719 & 0.793 & 0.797 & 0.909 \\
              &(0.266) &(0.218) &(0.23) &(0.187) &(0.097) &(0.09) \\ \hline
    joint     & 0.651 & 0.710 & 0.834 & 0.832 & 0.823 & 0.922 \\
              &(0.23) &(0.157) &(0.171) &(0.148) &(0.076) &(0.072) \\ \hline
    \end{tabular}
    \caption{Effect on the segmentation performance when the generator is optimized jointly with the segmentation networks' loss against the independent optimization of generator without segmentation loss. Mean Dice score with standard deviations in brackets is presented for cardiac dataset (see~\ref{disjoint_opti}-B).}
    %Effect of the segmentation loss in the generator optimization is presented on the cardiac dataset (mean DSC and standard deviations for $N_L=1$ and $3$). Disjoint model only optimizes the regularization term, including adversarial and large-deviation terms, and joint model optimizes the entire cost.}% on the segmentation performance when the generators are optimized without the segmentation loss(see~\ref{disjoint_opti}), presented for the cardiac dataset (mean dice score and standard deviation values for $N_{L}=$1 and 3).}
    \label{tab:non_joint_opti}
    %\vspace{-0.3cm}
\end{table}

%Varying number of unlabeled images
%\subsection{Varying number of unlabeled images}\label{vary_no_unl}
\textbf{C. Varying number of unlabeled images}\label{vary_no_unl}: We investigate how varying the number of unlabeled 3D volumes for the training of the proposed method can influence the segmentation performance.
This experiment is illustrated in Fig.~\ref{fig:vary_unlabeled_data} for the cardiac dataset. 
For extremely low values of unlabeled volumes of 1 and 3, we observe a drop in performance. While for a higher value of 50 volumes, we observed negligible improvements compared to using 25 volumes.
We observe that the improvements in segmentation accuracy do not change significantly for other values of unlabeled volumes from 5 to 25.
This indicates that we need some amount of unlabeled volumes to obtain stated performance gains, and beyond a certain number of unlabeled volumes we do not get extra benefits.
%varying the number of unlabeled 3D volumes used.

%width=0.8\textwidth
% \begin{figure}[t!]
%     \includegraphics[width=5cm,height=5cm,keepaspectratio,trim={0.2cm 0cm 0.2cm 0.2cm},clip]{images/seg_results/acdc/ra_0_vary_no_of_unl_imgs_tr1_dsc_bar_plot.png}\\
%     \caption{The number of 3D unlabeled images ($N_{UL}$) used for the training of the proposed method(see~\ref{vary_no_unl}) are varied for the cardiac dataset for $N_L=1$.}
%     \vspace{-0.1cm}
%     \label{fig:vary_unlabeled_data}
% \end{figure}

%Varying number of labeled images
%\subsection{Varying number of labeled images}\label{vary_no_lab}

\textbf{D. Varying number of labeled images}\label{vary_no_lab}: Here, we investigated how the performance gap between affine, random, and proposed augmentations varies as we increase the number of training volumes involved in the training. We observe that the performance gap between the augmentation approaches reduces as we increase the number of labeled training volumes as shown in Fig.~\ref{fig:vary_labeled_data}.
This is expected since as the labeled examples increase, the network sees larger number cases and gains robustness to variations present in these images.
%This is expected since as the labeled examples increase, the network sees larger number cases with different shape and intensity characteristics and gains robustness to these variations.
For the case of 40 3D training volumes, we see that the performance of affine augmentations is almost similar to the proposed augmentations.
We observe that the statistically significant improvements exist until 10 labeled examples. Also, we see that with the proposed model, segmentation accuracy using 10 labeled examples is similar to using 40 examples using random augmentations.
%One striking observation is that with the proposed model, segmentation accuracy using 10 labeled examples is similar to using 40 examples using random augmentations. 
%random deformation and intensity augmentation. 
% \begin{figure}[t!]
%     \includegraphics[width=0.9\textwidth,trim={0.2cm 0cm 0.2cm 0.2cm},clip]{images/seg_results/acdc/diff_no_of_tr_imgs_bar_plot.png}\\
%     \caption{The number of 3D labeled images ($N_{L}=1,3,5,10,15,40$) used for the training of the proposed augmentations against random augmentations is varied(see~\ref{vary_no_lab}), mean DSC with error bars showing one standard deviation are presented for the cardiac dataset.}
%     \vspace{-0.1cm}
%     \label{fig:vary_labeled_data}
% \end{figure}

%Varying the set of labeled, unlabeled, val, test images
%\subsection{Different set of train, validation, test, and unlabeled 3D volumes}\label{vary_val_set}
\textbf{E. Different set of train, validation, test, and unlabeled 3D volumes}\label{vary_val_set}: Lastly, we show that the results hold for any randomly chosen dataset split, and does not overfit to a specific set of validation images as illustrated in Table~\ref{tab:diff_tr_tst_sets}. 
%\setlength{\belowcaptionskip}{-15pt}
% random deformations + intensities (RD+RI) & 0.506 +/- 0.213  & 0.647 +/- 0.159 & 0.819 +/- 0.149 \\ \hline
%     generated deformations + intensities (GD+GI) & 0.683 +/- 0.212 & 0.693 +/- 0.139 & 0.842 +/- 0.136 \\ \hline
\begin{table}[!t]
    \centering
%    \begin{tabular}{|p||||}
    \begin{tabular}{|p{1.5cm}|p{1.9cm}|p{1.9cm}|p{1.9cm}|}
    \hline
    %\multirow{2}{*}{Methods}  & \multicolumn{3}{c|}{Number of 3D training volumes used = 1}
    %\\ 
    %\multirow{2}{*}{ } & \multicolumn{3}{c|}{volumes used = 1}
    %\multirow{1}{*}{Methods} & RV & Myo & LV \\ \hline
    Methods & RV & Myo & LV \\ \hline
    RD+RI & 0.506 (0.213) & 0.647 (0.159) & 0.819 (0.149) \\ \hline
    GD+GI & 0.683 (0.212) & 0.693 (0.139) & 0.842 (0.136) \\ \hline
    \end{tabular}
    \caption{Mean Dice scores with standard deviations for the cardiac dataset for a different set of train, validation, test, and unlabeled volumes for $N_L=1$ (see~\ref{vary_val_set}-E).}
    \label{tab:diff_tr_tst_sets}
    %\vspace{-0.3cm}
\end{table}

%No validation images
%\subsection{No validation images}\label{no_val_set}
\textbf{F. No validation images}\label{no_val_set}: Additionally, we also present results for the case when no validation images were used in the training in Fig.~\ref{fig:no_val_imgs} in the appendix. Here, the chosen model parameters are obtained after training the generators for predefined number of iterations. Surprisingly, we observe that the performance obtained without validation images does not vary much w.r.t the performance obtained using validation images. %(shown in Figure~\ref{fig:no_val_imgs} in Appendix) 

% ================================================================
% qualitative results - seg masks
% ================================================================
%\setlength{\belowcaptionskip}{-25pt}
%\setlength{\intextsep}{-10pt}

% ================================================================
% qualitative results - cganDeform
% ================================================================
%\setlength{\belowcaptionskip}{-25pt}
%\setlength{\intextsep}{-10pt}
%\vspace{-0.2cm}

\section{Conclusion}
In the clinical setting, deployment of successful deep learning algorithms for medical image analysis is limited due to the difficulty of assembling large-scale annotated datasets. In this work, we proposed a semi-supervised task-driven data augmentation method to tackle the issue of obtaining robust segmentation in limited data setting for training. To achieve this we proposed two novel contributions: (i) task-driven based optimization where the generation of the augmentation data is optimal for the segmentation performance, and (ii) semi-supervised nature is induced by using the unlabeled data in the generative modeling setup, where we design two conditional generative models to output transformations that capture two factors of variations: shape and intensity characteristics present in the population. Using three publicly available datasets, we demonstrated the proposed method for segmenting the cardiac, prostate and pancreas using limited annotated examples, reporting substantial performance gains over existing methods. Surprisingly, the augmented images generated via the proposed task-driven approach were not necessarily realistic yet yielded improved segmentation performance, questioning the validity of the assumption that generating realistic examples is the optimal way. 
\vspace{-0.2cm}

\section{Acknowledgements}

% The presented work is partially funded by: 
% 1. Swiss Data Science Center (DeepMicroIA), 
% 2. Clinical Research Priority Program Grant (CRPP) on Artificial Intelligence in Oncological Imaging Network, University of Zurich, 
% 3. Swiss Platform for Advanced Scientific Computing (PASC), coordinated by Swiss National Super-computing Centre (CSCS), 
% 4. Personalized Health and Related Technologies (PHRT), project number 222, ETH domain. 
% We thank Nvidia for their GPU donation.
The presented work is partially funded by: 
1. Swiss Data Science Center (DeepMicroIA), 
2. Clinical Research Priority Program Grant (CRPP) on Artificial Intelligence in Oncological Imaging Network, University of Zurich, 
3. Platform for Advanced Scientific Computing (PASC) - project HCP-Predict,
4. Personalized Health Related Technologies - project number 222, ETH.
%Swiss Platform for Advanced Scientific Computing (PASC), coordinated by Swiss National Super-computing Centre (CSCS), 
%4. Personalized Health and Related Technologies (PHRT), project number 222, ETH domain. 
We thank Nvidia for their GPU donation.

%\section{References}

% Please ensure that every reference cited in the text is also present in
% the reference list (and vice versa).

% \section*{\itshape Reference style}

% Text: All citations in the text should refer to:
% \begin{enumerate}
% \item Single author: the author's name (without initials, unless there
% is ambiguity) and the year of publication;
% \item Two authors: both authors' names and the year of publication;
% \item Three or more authors: first author's name followed by `et al.'
% and the year of publication.
% \end{enumerate}
% Citations may be made directly (or parenthetically). Groups of
% references should be listed first alphabetically, then chronologically.

%%Harvard
%\bibliographystyle{model2-names.bst}\biboptions{authoryear}
\bibliographystyle{IEEEtran}
\bibliography{references.bib}

\newpage
%\section*{Supplementary Material}
\section{Appendix}

\begin{figure}[!ht]
    \includegraphics[width=0.9\textwidth,trim={0.2cm 0cm 0.2cm 0.2cm},clip]{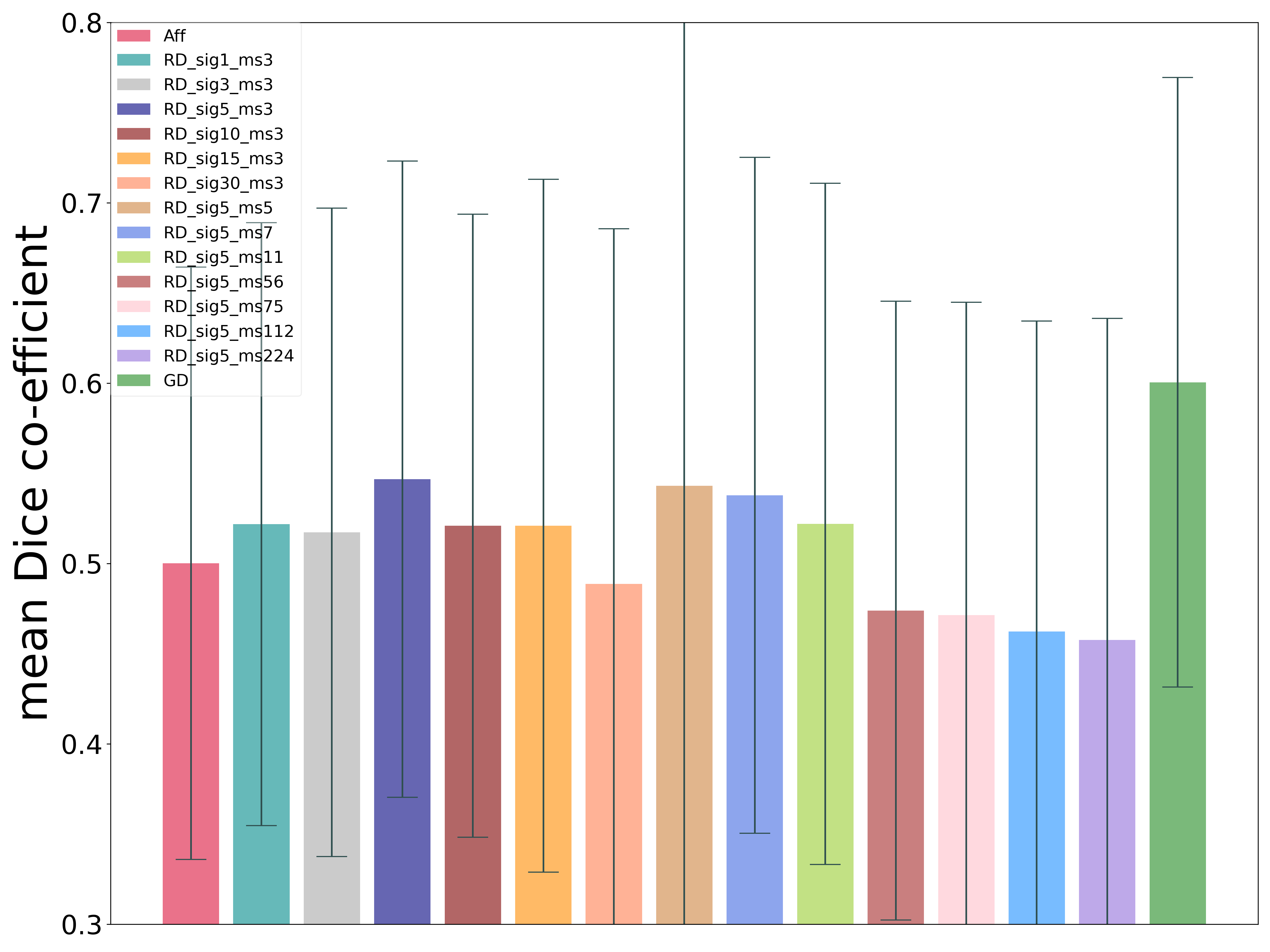}\\
    \caption{The segmentation performance was obtained using random elastic augmented images (RD). These RD images are generated using elastic deformation fields as defined in~\cite{ronneberger2015u} applied to the input image. We evaluate different initial matrix sizes (ms x ms x 2) and sigma values that are later interpolated to image dimensions to obtain the deformation fields. These matrix values are randomly sampled from a Gaussian distribution of zero mean and standard deviation of sigma.
    Here, GD denotes that the augmented images are generated using the learned deformation field generator $G_V$.
    (mean Dice scores are reported for $N_L=1$ for the cardiac dataset).}
    \vspace{-0.1cm}
    \label{fig:rd_vary_sig_ks}
\end{figure}

\begin{figure}[!ht]
    \includegraphics[width=0.9\textwidth,trim={0.2cm 0cm 0.2cm 0.2cm},clip]{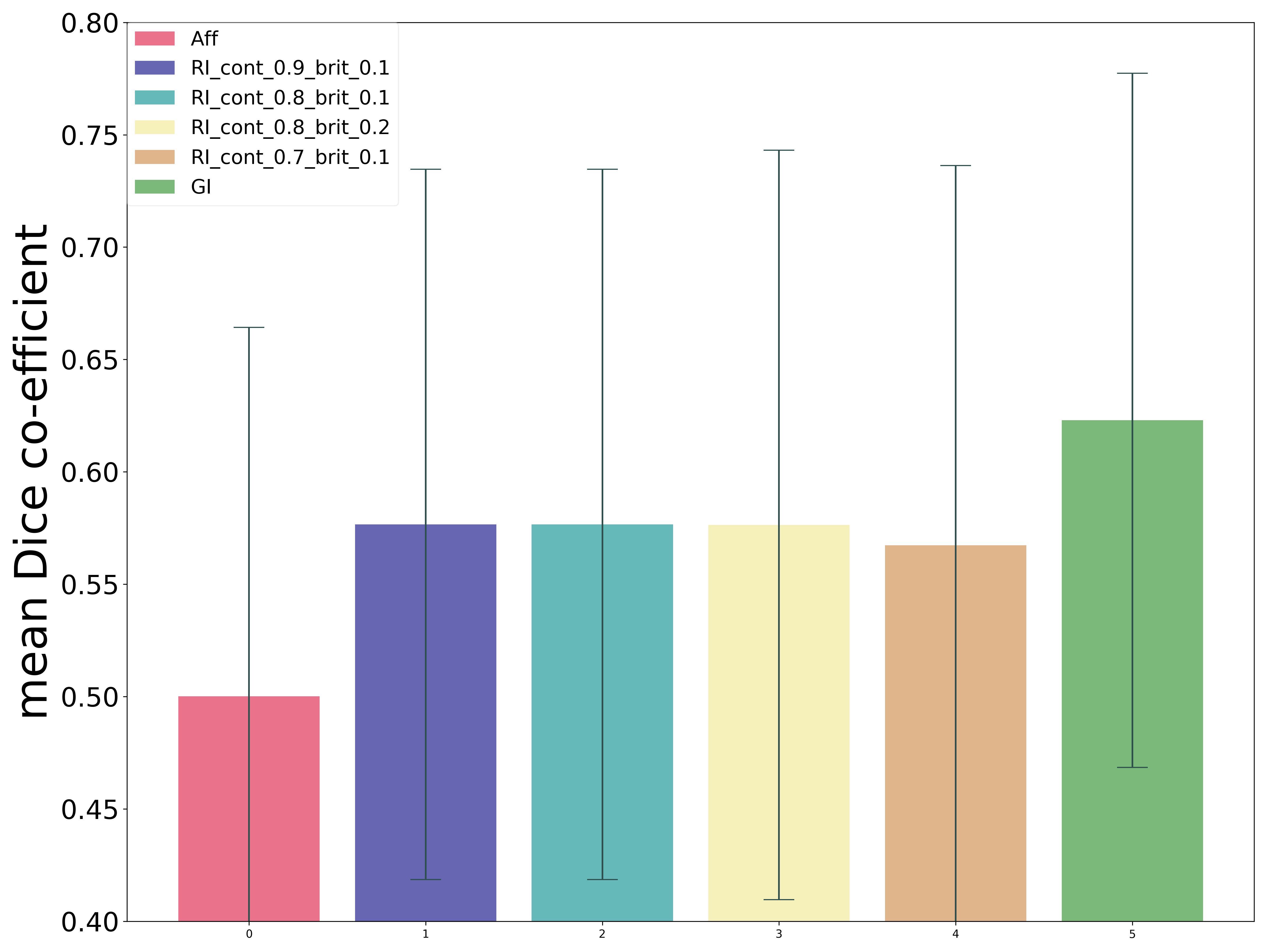}\\
    \caption{The segmentation performance obtained using different combinations of contrast (cont) and brightness (brit) values used to generate intensity transformation fields as defined in~\cite{hong2017convolutional,perez2018data}. Here, GI denotes that the augmented images are generated using the learned intensity field generator $G_I$. (mean Dice scores are reported for $N_L=1$ for the cardiac dataset).}
    \vspace{-0.1cm}
    \label{fig:rd_vary_cont_brit}
\end{figure}

\begin{figure}[!ht]
    \includegraphics[width=0.9\textwidth,trim={0.2cm 0cm 0.2cm 0.2cm},clip]{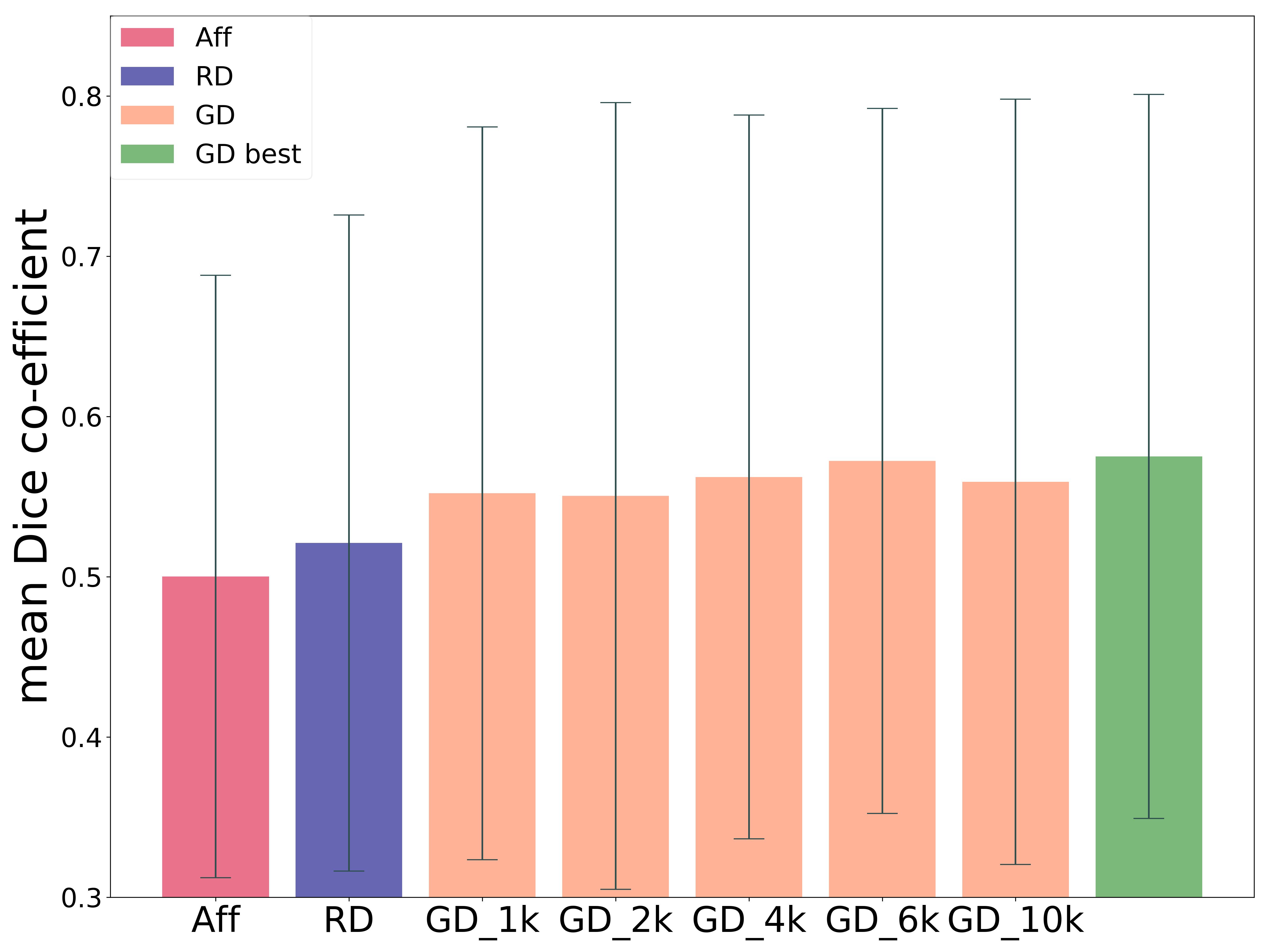}\\
    \caption{The segmentation performance obtained for the case of using deformation field generator (GD) when no validation images are used during the training of the generator, and the training is stopped after a pre-defined number of training iterations.
    Here, $\_xk$ denotes the pre-defined number of training iterations, where $\_1k,\_2k,\_4k,\_6k,\_10k$ denotes that the model is trained for 1000, 2000, 4000, 6000, 10000 iterations, respectively. 
    $\_best$ denotes the case where two 3D validations volumes are used to determine the model parameters based on best validation Dice score observed through all the 10000 training iterations. We additionally compare it against random deformations (RD) (mean Dice scores is presented for $N_L=1$ for the cardiac dataset).}
    \vspace{-0.1cm}
    \label{fig:no_val_imgs}
\end{figure}

\begin{table}[!hb]
\begin{center}
\begin{tabular}{|p{2.5cm}|p{1.5cm}|p{1.5cm}|p{1.5cm}|}
\hline
{method} & {RV } & {Myo } & {LV } \\ \hline
{random deformations $+$ intensities (RD $+$ RI)} & 0.241 (0.224) & 0.368 (0.246) & 0.44 (0.296) \\ \hline
{learned deformations $+$ intensities (GD $+$ GI)} & 0.411 (0.28) $\star$  & 0.587 (0.248) $\star$ & 0.644 (0.249) $\star$ \\ \hline
\end{tabular}
\caption{ACDC dataset - mean DSC (std dev) over 15 runs is reported for 20 test \textbf{ED} images when segmentation network was trained with 1 labeled \textbf{ED} volume with the proposed learned augmented images (GD$+$GI) which is compared to random augmentations (RD$+$RI) ($\star$ denotes the statistical significance of the proposed method (GD $+$ GI) over random augmentations (RD $+$ RI) illustrated using the Wilcoxon signed-rank test with a threshold p-value of 0.05).}
\label{table:acdc_ed_images}
\end{center}
\end{table}

\end{document}